\documentclass[a4paper, 11pt]{article}
\pdfoutput=1

\usepackage[utf8]{inputenc}
\usepackage{jheppub}
\usepackage{afterpage}
\usepackage{ifthen}

\newcommand{\Dmq}{\Delta m^2}
\newcommand{\eVq}{\ensuremath{\text{eV}^2}}

\renewcommand{\Im}{\mathop{\rm Im}}
\newenvironment{pagefigure}{\begin{figure}[!p]}{\afterpage\clearpage\end{figure}}

\title{Updated fit to three neutrino mixing: exploring the
  accelerator--reactor complementarity}

\author[a]{Ivan Esteban,}
\affiliation[a] {Departament de Fis\'{\i}ca Qu\`antica i
  Astrof\'{\i}sica and Institut de Ciencies del Cosmos, Universitat de
  Barcelona, Diagonal 647, E-08028 Barcelona, Spain}
\emailAdd{ivan.esteban@fqa.ub.edu}

\author[a,b,c]{M.~C.~Gonzalez-Garcia,}
\affiliation[b]{Instituci\'o Catalana de Recerca i Estudis
  Avan\c{c}ats (ICREA), Pg. Lluis Companys 23, 08010 Barcelona,
  Spain.}
\affiliation[c]{C.N.~Yang Institute for Theoretical Physics, State
  University of New York at Stony Brook, Stony Brook, NY 11794-3840,
  USA}
\emailAdd{maria.gonzalez-garcia@stonybrook.edu}

\author[d]{Michele Maltoni,}
\affiliation[d]{Instituto de F\'{\i}sica Te\'orica UAM/CSIC, Calle de
  Nicol\'as Cabrera 13--15, Universidad Aut\'onoma de Madrid,
  Cantoblanco, E-28049 Madrid, Spain}
\emailAdd{michele.maltoni@csic.es}

\author[d]{Ivan Martinez-Soler,}
\emailAdd{ivanj.m@csic.es}

\author[e]{Thomas Schwetz}
\affiliation[e]{Institut f\"ur Kernphysik, Karlsruher Institut f\"ur
  Technologie (KIT), D-76021 Karlsruhe, Germany}
\emailAdd{schwetz@kit.edu}

\abstract{We perform a combined fit to global neutrino oscillation
  data available as of fall 2016 in the scenario of three-neutrino
  oscillations and present updated allowed ranges of the six
  oscillation parameters. We discuss the differences arising between
  the consistent combination of the data samples from accelerator and
  reactor experiments compared to partial combinations. We quantify
  the confidence in the determination of the less precisely known
  parameters $\theta_{23}$, $\delta_\text{CP}$, and the neutrino mass
  ordering by performing a Monte Carlo study of the long baseline
  accelerator and reactor data.  We find that the sensitivity to the
  mass ordering and the $\theta_{23}$ octant is below
  $1\sigma$. Maximal $\theta_{23}$ mixing is allowed at slightly more
  than 90\%~CL. The best fit for the CP violating phase is around
  $270^\circ$, CP conservation is allowed at slightly above $1\sigma$,
  and values of $\delta_\text{CP} \simeq 90^\circ$ are disfavored at
  around 99\%~CL for normal ordering and higher CL for inverted
  ordering.}

\preprint{IFT-UAM/CSIC-16-114, YITP-SB-16-45}


\begin{document}

\maketitle

\section{Introduction}

Experiments measuring the flavor composition of solar neutrinos,
atmospheric neutrinos, neutrinos produced in nuclear reactors and in
accelerators have established that lepton flavor is not conserved in
neutrino propagation, but it oscillates with a wavelength depending on
distance and energy, because neutrinos are massive and the mass states
are admixtures of the flavor states~\cite{Pontecorvo:1967fh,
  Gribov:1968kq}, see Ref.~\cite{GonzalezGarcia:2007ib} for an
overview.

With the exception of a set of unconfirmed ``hints'' of possible eV
scale mass states (see Ref.~\cite{Giunti:2015wnd} for a recent
review), all the oscillation signatures can be explained with the
three flavor neutrinos ($\nu_e$, $\nu_\mu$, $\nu_\tau$), which can be
expressed as quantum superpositions of three massive states $\nu_i$
($i=1,2,3$) with masses $m_i$.  This implies the presence of a
leptonic mixing matrix in the weak charged current
interactions~\cite{Maki:1962mu, Kobayashi:1973fv} which can be
parametrized as:
\begin{equation}
  \label{eq:U3m}
  U =
  \begin{pmatrix}
    1 & 0 & 0 \\
    0 & c_{23}  & {s_{23}} \\
    0 & -s_{23} & {c_{23}}
  \end{pmatrix}
  \cdot
  \begin{pmatrix}
    c_{13} & 0 & s_{13} e^{-i\delta_\text{CP}} \\
    0 & 1 & 0 \\
    -s_{13} e^{i\delta_\text{CP}} & 0 & c_{13}
  \end{pmatrix}
  \cdot
  \begin{pmatrix}
    c_{12} & s_{12} & 0 \\
    -s_{12} & c_{12} & 0 \\
    0 & 0 & 1
  \end{pmatrix}
  \cdot \mathcal{P}
\end{equation}
where $c_{ij} \equiv \cos\theta_{ij}$ and $s_{ij} \equiv
\sin\theta_{ij}$.  The angles $\theta_{ij}$ can be taken without loss
of generality to lie in the first quadrant, $\theta_{ij} \in [0,
  \pi/2]$, and the phase $\delta_\text{CP} \in [0, 2\pi]$.  Here
$\mathcal{P}$ is a diagonal matrix which is the identity if neutrinos
are Dirac fermions and it contains two additional phases if they are
Majorana fermions, and plays no role in neutrino
oscillations~\cite{Bilenky:1980cx, Langacker:1986jv}.
In this convention there are two non-equivalent orderings for the
neutrino masses which can be chosen to be: normal ordering (NO) with
$m_1 < m_2 < m_3$, and inverted ordering (IO) with $m_3 < m_1 <
m_2$. Furthermore the data shows a relatively large hierarchy between
the mass splittings, $\Dmq_{21} \ll |\Dmq_{31}| \simeq |\Dmq_{32}|$
with $\Dmq_{ij} \equiv m_i^2 - m_j^2$.  In this work we follow the
convention introduced in Ref.~\cite{Gonzalez-Garcia:2014bfa} and
present our results in terms of the variable $\Dmq_{3\ell}$, with
$\ell=1$ for NO and $\ell=2$ for IO.  Hence, $\Dmq_{3\ell} = \Dmq_{31}
> 0$ for NO and $\Dmq_{3\ell} = \Dmq_{32} < 0$ for IO, \textit{i.e.},
it corresponds to the mass splitting with the largest absolute value.

In this article, we present an up-to-date (as of fall 2016) global
analysis of neutrino data in the framework of three-neutrino
oscillations. Alternative recent global fits have been presented in
Refs.~\cite{Capozzi:2016rtj, Forero:2014bxa}. With current data from
the accelerator long-baseline experiments MINOS, T2K, NO$\nu$A and
modern reactor experiments like Daya-Bay, RENO, and Double-Chooz,
their complementarity anticipated more than a decade
ago~\cite{Minakata:2002jv, Huber:2003pm, Huber:2004ug} has become a
reality, and the combined analysis starts to show some sensitivity to
subtle effects like the $\theta_{23}$ octant or the $\delta_\text{CP}$
phase (though still at low statistical significance).

The outline of the paper is as follows: In Sec.~\ref{subsec:data} we
describe the data samples included in our analysis (see also
Appendix~\ref{sec:appendix} for a schematic list).  The presently
allowed ranges of the six oscillation parameters are given in
Sec.~\ref{subsec:oscparam} assuming that $\Delta\chi^2$ follows a
$\chi^2$-distribution, while Sec.~\ref{subsec:CP} contains the
corresponding measures of CP violation in terms of the leptonic
Jarlskog invariant and the leptonic unitarity triangle.  Deviations
from the Gaussian approximation of the confidence intervals for
$\theta_{23}$ and $\delta_\text{CP}$ and the confidence level for the
mass ordering determination are quantified in Sec.~\ref{sec:MC}.
Several issues appearing in the present analysis are discussed in
Sec.~\ref{sec:issues}, in particular about the consistent combination
of results from long baseline accelerator experiments with reactors
results, now that both provide comparable precision in the
determination of the relevant mass-squared difference. We also give
the updated status on the ongoing tension in the $\Dmq_{21}$
determination from solar experiments versus KamLAND, and comment on
the stand-by in the analysis of the Super-Kamiokande atmospheric
data. Sec.~\ref{sec:summary} contains the summary of our results.

\section{Global analysis: determination of oscillation parameters}
\label{sec:global}

\subsection{Data samples analyzed}
\label{subsec:data}

In the analysis of solar neutrino data we consider the total rates
from the radiochemical experiments Chlorine~\cite{Cleveland:1998nv},
Gallex/GNO~\cite{Kaether:2010ag} and SAGE~\cite{Abdurashitov:2009tn},
the results for the four phases of
Super-Kamiokande~\cite{Hosaka:2005um, Cravens:2008aa, Abe:2010hy,
  sksol:nakano2016, sksol:ichep2016}, the data of the three phases of
SNO included in the form of the parametrization presented
in~\cite{Aharmim:2011vm}, and the results of both Phase-I and Phase-II
of Borexino~\cite{Bellini:2011rx, Bellini:2008mr, Bellini:2014uqa}.

Results from long baseline (LBL) accelerator experiments include the
final energy distribution of events from MINOS~\cite{Adamson:2013whj,
  Adamson:2013ue} in $\nu_\mu$ and $\bar\nu_\mu$ disappearance and
$\nu_e$ and $\bar\nu_e$ appearance channels, as well as the latest
energy spectrum for T2K in the same four channels~\cite{t2k:ichep2016,
  t2k:susy2016} and for NO$\nu$A on the $\nu_\mu$ disappearance and
$\nu_e$ appearance neutrino modes~\cite{nova:nu2016}.

Data samples on $\bar\nu_e$ disappearance from reactor include the full
results of the long baseline reactor data in
KamLAND~\cite{Gando:2010aa}, as well as the results from medium
baseline reactor experiments from CHOOZ~\cite{Apollonio:1999ae} and
Palo Verde~\cite{Piepke:2002ju}. Concerning running experiments we
include the latest spectral data from
Double-Chooz~\cite{dc:moriond2016} and Daya-Bay~\cite{db:nu2016},
while for RENO we use the total rates obtained with their largest data
sample corresponding to 800 days of data-taking~\cite{reno:nu2014}.

In the analysis of the reactor data, the unoscillated reactor flux is
determined as described in~\cite{Kopp:2013vaa} by including in the fit
the results from short baseline reactor data (RSBL) from
ILL~\cite{Kwon:1981ua}, G\"osgen~\cite{Zacek:1986cu},
Krasnoyarsk~\cite{Vidyakin:1987ue, Vidyakin:1994ut},
ROVNO88~\cite{Afonin:1988gx}, ROVNO4~\cite{Kuvshinnikov:1990ry},
Bugey3~\cite{Declais:1994su}, Bugey4~\cite{Declais:1994ma}, and
SRP~\cite{Greenwood:1996pb}.

For the analysis of atmospheric neutrinos we include the results from
IceCube/DeepCore 3-year data~\cite{Aartsen:2014yll}.

The above data sets constitute the samples included in our NuFIT 3.0
analysis.  For Super-Kamiokande atmospheric neutrino data from phases
SK1--4 we will comment on our strategy in Sec.~\ref{subsec:SK}.  A
full list of experiments including the counting of data points in each
sample can be found in Appendix~\ref{sec:appendix}.

\subsection{Results: oscillation parameters}
\label{subsec:oscparam}

The results of our standard analysis are presented in
Figs.~\ref{fig:region-glob} and~\ref{fig:chisq-glob} where we show
projections of the allowed six-dimensional parameter
space.\footnote{$\Delta\chi^2$ tables from the global analysis
  corresponding to all 1-dimensional and 2-dimensional projections are
  available for download at the NuFIT website~\cite{nufit}.}  In all
cases when including reactor experiments we leave the normalization of
reactor fluxes free and include data from short-baseline (less than
100 m) reactor experiments. In our previous
analysis~\cite{Gonzalez-Garcia:2014bfa, GonzalezGarcia:2012sz} we
studied the impact of this choice versus that of fixing the reactor
fluxes to the prediction of the latest
calculations~\cite{Mueller:2011nm, Huber:2011wv, Mention:2011rk}.  As
expected, the overall description is better when the flux
normalization $f_\text{flux}$ is fitted against the data.  We find
$\chi^2(f_\text{flux}~\text{fix}) - \chi^2(f_\text{flux}~\text{fit})
\simeq 6$ which is just another way to quantify the well-known short
baseline reactor anomaly to be $\sim 2.5\sigma$.  However, the
difference in the resulting parameter determination (in particular for
$\theta_{13}$) between these two reactor flux normalization choices
has become marginal, since data from the reactor experiments with near
detectors such as Daya-Bay, RENO and Double-Chooz (for which the
near-far comparison allows for flux-normalization independent
analysis) is now dominant.  Consequently, in what follows we show only
the $\Delta\chi^2$ projections for our standard choice with fitted
reactor flux normalization.

\begin{pagefigure}\centering
  \includegraphics[width=0.81\textwidth]{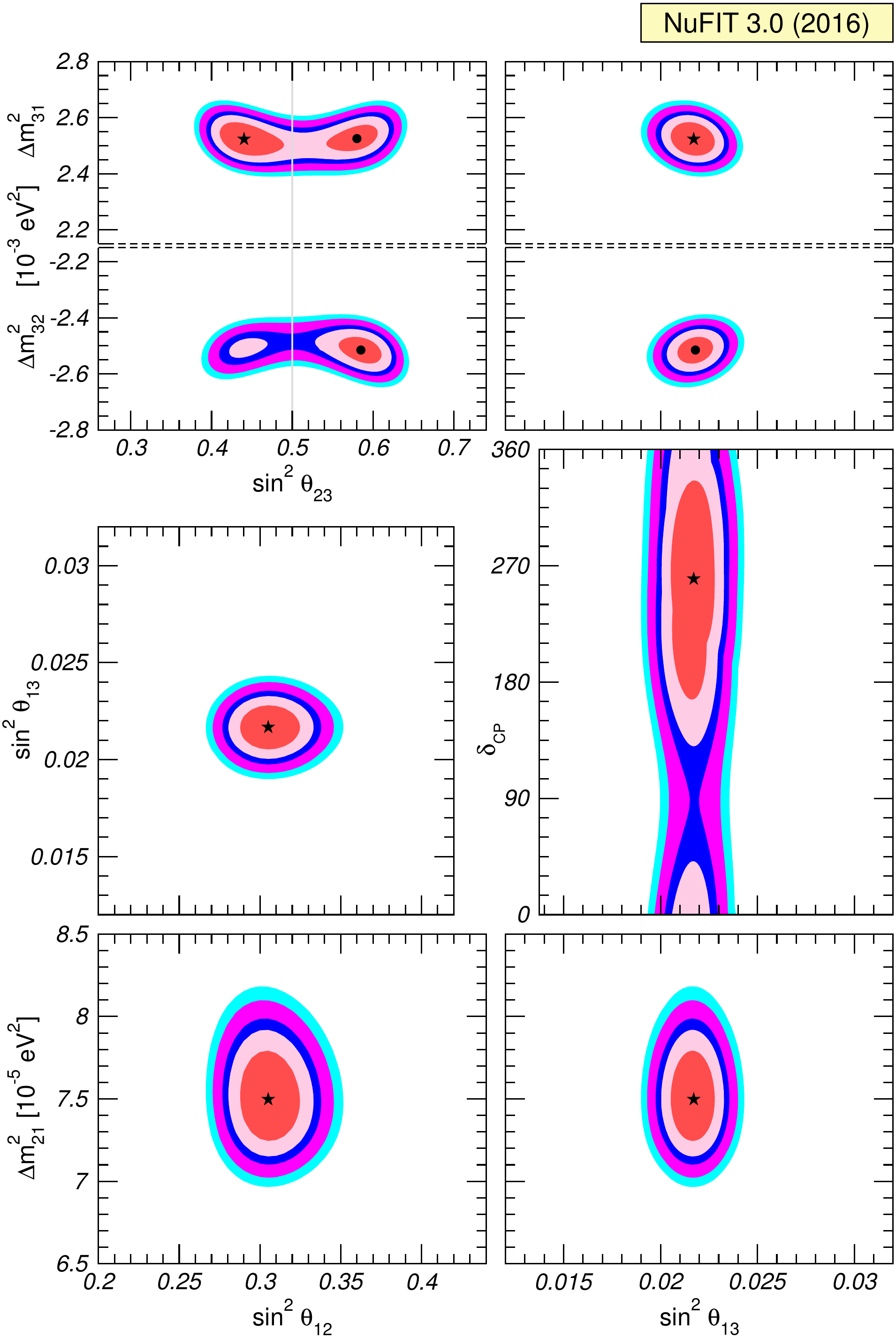}
  \caption{Global $3\nu$ oscillation analysis. Each panel shows the
    two-dimensional projection of the allowed six-dimensional region
    after marginalization with respect to the undisplayed
    parameters. The different contours correspond to $1\sigma$, 90\%,
    $2\sigma$, 99\%, $3\sigma$ CL (2 dof).  The normalization of
    reactor fluxes is left free and data from short-baseline reactor
    experiments are included as explained in the text. Note that as
    atmospheric mass-squared splitting we use $\Dmq_{31}$ for NO and
    $\Dmq_{32}$ for IO. The regions in the four lower panels are
    obtained from $\Delta\chi^2$ minimized with respect to the mass
    ordering.}
  \label{fig:region-glob}
\end{pagefigure}

\begin{pagefigure}\centering
\includegraphics[width=0.86\textwidth]{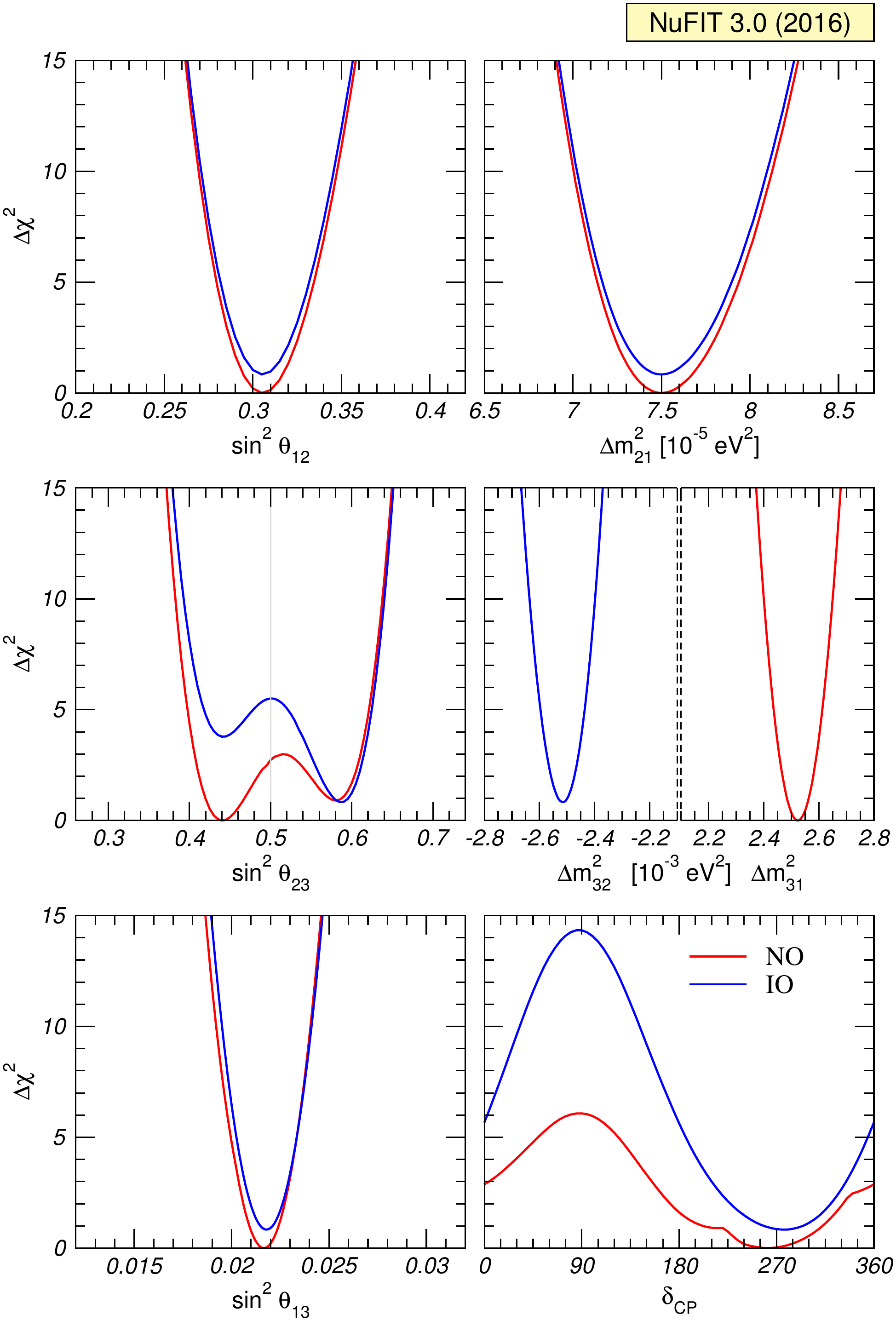}
  \caption{Global $3\nu$ oscillation analysis. The red (blue) curves
    correspond to Normal (Inverted) Ordering.  The normalization of
    reactor fluxes is left free and data from short-baseline reactor
    experiments are included.  Note that as atmospheric mass-squared
    splitting we use $\Dmq_{31}$ for NO and $\Dmq_{32}$ for IO.}
  \label{fig:chisq-glob}
\end{pagefigure}

The best fit values and the derived ranges for the six parameters at
the $1\sigma$ ($3\sigma$) level are given in Tab.~\ref{tab:bfranges}.
For each parameter $x$ the ranges are obtained after marginalizing
with respect to the other parameters\footnote{In this paper we use the
  term ``marginalization'' over a given parameter as synonym for
  minimizing the $\chi^2$ function with respect to that parameter.}
and under the assumption that $\Delta\chi_\text{marg}^2(x)$ follows a
$\chi^2$ distribution. Hence the $1\sigma$ ($3\sigma$) ranges are
given by the condition $\Delta\chi_\text{marg}^2(x)=1$ (9). It is
known that because of its periodic nature and the presence of
parameter degeneracies the statistical distribution of the
marginalized $\Delta\chi^2$ for $\delta_\text{CP}$ and $\theta_{23}$
(and consequently the corresponding CL intervals) may be
modified~\cite{Schwetz:2006md, Blennow:2014sja}.  In Sec.~\ref{sec:MC}
we will discuss and quantify these effects.

In Tab.~\ref{tab:bfranges} we list the results for three scenarios. In
the first and second columns we assume that the ordering of the
neutrino mass states is known \emph{a priori} to be Normal or
Inverted, respectively, so the ranges of all parameters are defined
with respect to the minimum in the given scenario.  In the third
column we make no assumptions on the ordering, so in this case the
ranges of the parameters are defined with respect to the global
minimum (which corresponds to Normal Ordering) and are obtained
marginalizing also over the ordering. For this third case we only give
the $3\sigma$ ranges. In this case the range of $\Dmq_{3\ell}$ is
composed of two disconnected intervals, one containing the absolute
minimum (NO) and the other the secondary local minimum (IO).

\begin{table}\centering
  \begin{footnotesize}
    \begin{tabular}{l|cc|cc|c}
      \hline\hline
      & \multicolumn{2}{c|}{Normal Ordering (best fit)}
      & \multicolumn{2}{c|}{Inverted Ordering ($\Delta\chi^2=0.83$)}
      & Any Ordering
      \\
      \hline
      & bfp $\pm 1\sigma$ & $3\sigma$ range
      & bfp $\pm 1\sigma$ & $3\sigma$ range
      & $3\sigma$ range
      \\
      \hline
      \rule{0pt}{4mm}\ignorespaces
      $\sin^2\theta_{12}$
      & $0.306_{-0.012}^{+0.012}$ & $0.271 \to 0.345$
      & $0.306_{-0.012}^{+0.012}$ & $0.271 \to 0.345$
      & $0.271 \to 0.345$
      \\[1mm]
      $\theta_{12}/^\circ$
      & $33.56_{-0.75}^{+0.77}$ & $31.38 \to 35.99$
      & $33.56_{-0.75}^{+0.77}$ & $31.38 \to 35.99$
      & $31.38 \to 35.99$
      \\[3mm]
      $\sin^2\theta_{23}$
      & $0.441_{-0.021}^{+0.027}$ & $0.385 \to 0.635$
      & $0.587_{-0.024}^{+0.020}$ & $0.393 \to 0.640$
      & $0.385 \to 0.638$
      \\[1mm]
      $\theta_{23}/^\circ$
      & $41.6_{-1.2}^{+1.5}$ & $38.4 \to 52.8$
      & $50.0_{-1.4}^{+1.1}$ & $38.8 \to 53.1$
      & $38.4 \to 53.0$
      \\[3mm]
      $\sin^2\theta_{13}$
      & $0.02166_{-0.00075}^{+0.00075}$ & $0.01934 \to 0.02392$
      & $0.02179_{-0.00076}^{+0.00076}$ & $0.01953 \to 0.02408$
      & $0.01934 \to 0.02397$
      \\[1mm]
      $\theta_{13}/^\circ$
      & $8.46_{-0.15}^{+0.15}$ & $7.99 \to 8.90$
      & $8.49_{-0.15}^{+0.15}$ & $8.03 \to 8.93$
      & $7.99 \to 8.91$
      \\[3mm]
      $\delta_\text{CP}/^\circ$
      & $261_{-59}^{+51}$ & $\hphantom{00}0 \to 360$
      & $277_{-46}^{+40}$ & $145 \to 391$
      & $\hphantom{00}0 \to 360$
      \\[3mm]
      $\dfrac{\Dmq_{21}}{10^{-5}~\eVq}$
      & $7.50_{-0.17}^{+0.19}$ & $7.03 \to 8.09$
      & $7.50_{-0.17}^{+0.19}$ & $7.03 \to 8.09$
      & $7.03 \to 8.09$
      \\[3mm]
      $\dfrac{\Dmq_{3\ell}}{10^{-3}~\eVq}$
      & $+2.524_{-0.040}^{+0.039}$ & $+2.407 \to +2.643$
      & $-2.514_{-0.041}^{+0.038}$ & $-2.635 \to -2.399$
      & $\begin{bmatrix}
        +2.407 \to +2.643\\[-2pt]
        -2.629 \to -2.405
      \end{bmatrix}$
      \\[3mm]
      \hline\hline
    \end{tabular}
  \end{footnotesize}
  \caption{Three-flavor oscillation parameters from our fit to global
    data after the NOW~2016 and ICHEP-2016 conference.  The numbers in
    the 1st (2nd) column are obtained assuming NO (IO), \textit{i.e.},
    relative to the respective local minimum, whereas in the 3rd
    column we minimize also with respect to the ordering. Note that
    $\Dmq_{3\ell} \equiv \Dmq_{31} > 0$ for NO and $\Dmq_{3\ell}
    \equiv \Dmq_{32} < 0$ for IO.}
  \label{tab:bfranges}
\end{table}

Defining the $3\sigma$ relative precision of a parameter by
$2(x^\text{up} - x^\text{low}) / (x^\text{up} + x^\text{low})$, where
$x^\text{up}$ ($x^\text{low}$) is the upper (lower) bound on a
parameter $x$ at the $3\sigma$ level, we read $3\sigma$ relative
precision of 14\% ($\theta_{12}$), 32\% ($\theta_{23}$), 11\%
($\theta_{13}$), 14\% ($\Dmq_{21}$) and 9\% ($|\Dmq_{3\ell}|$) for the
various oscillation parameters.

\subsection{Results: leptonic mixing matrix and CP violation}
\label{subsec:CP}

From the global $\chi^2$ analysis described in the previous section
and following the procedure outlined in
Ref.~\cite{GonzalezGarcia:2003qf} one can derive the $3\sigma$ ranges
on the magnitude of the elements of the leptonic mixing matrix:
\begin{equation}
  \label{eq:umatrix}
  |U| = \begin{pmatrix}
    0.800 \to 0.844 &\qquad
    0.515 \to 0.581 &\qquad
    0.139 \to 0.155
    \\
    0.229 \to 0.516 &\qquad
    0.438 \to 0.699 &\qquad
    0.614 \to 0.790
    \\
    0.249 \to 0.528 &\qquad
    0.462 \to 0.715 &\qquad
    0.595 \to 0.776
  \end{pmatrix} .
\end{equation}
Note that there are strong correlations between the elements due to
the unitary constraint.

\begin{figure}\centering
  \includegraphics[width=0.9\textwidth]{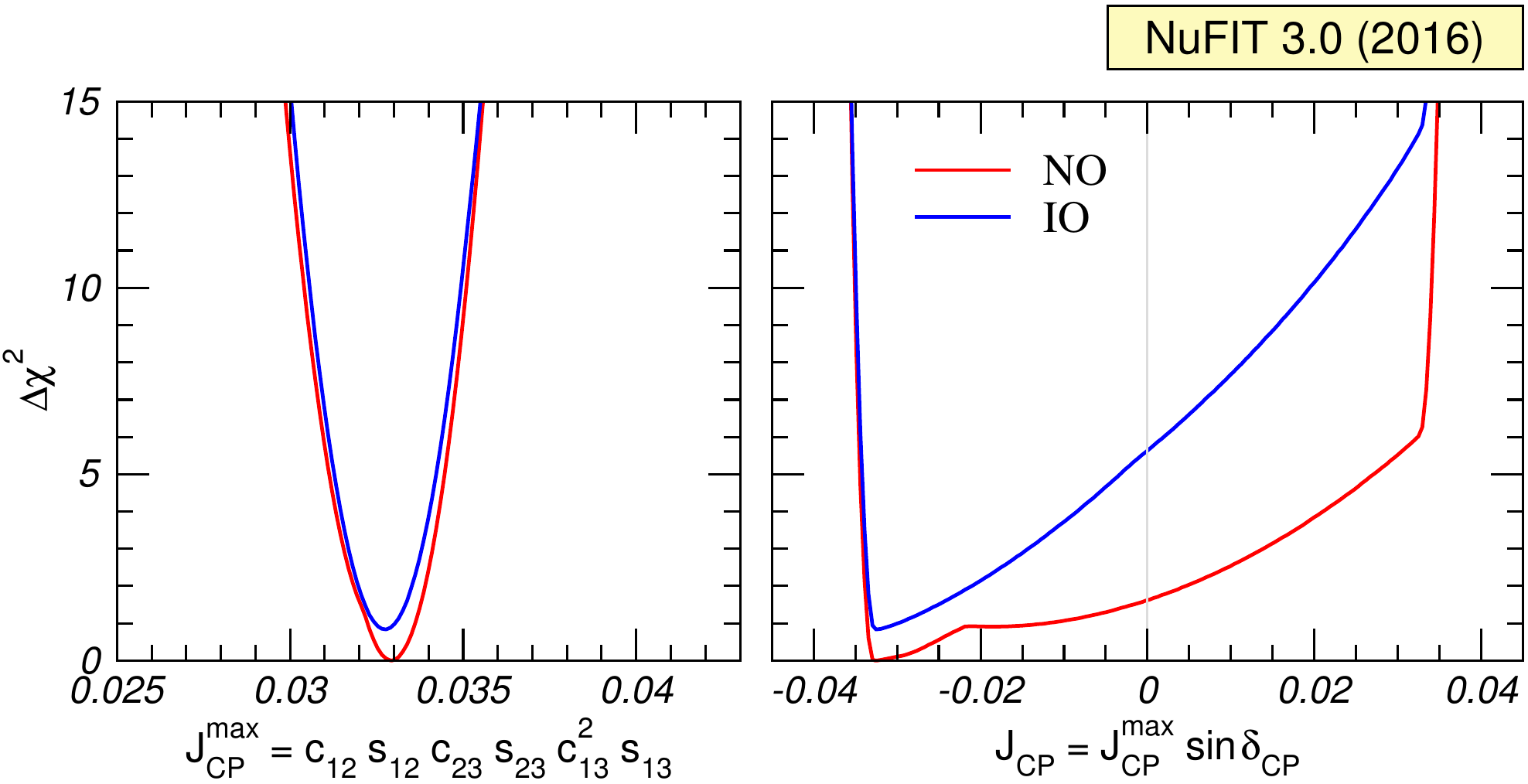}
  \caption{Dependence of the global $\Delta\chi^2$ function on the
    Jarlskog invariant. The red (blue) curves are for NO (IO).}
  \label{fig:chisq-viola}
\end{figure}

The present status of the determination of leptonic CP violation is
illustrated in Fig.~\ref{fig:chisq-viola}. In the left panel we show
the dependence of $\Delta\chi^2$ of the global analysis on the
Jarlskog invariant which gives a convention-independent measure of CP
violation~\cite{Jarlskog:1985ht}, defined as usual by:
\begin{equation}
  \Im\big[ U_{\alpha i} U_{\alpha j}^* U_{\beta i}^* U_{\beta j} \big]
  \equiv J_\text{CP}^\text{max} \sin\delta =
  \cos\theta_{12} \sin\theta_{12}
  \cos\theta_{23} \sin\theta_{23} \cos^2\theta_{13} \sin\theta_{13}
  \sin\delta
\end{equation}
where we have used the parametrization in Eq.~\eqref{eq:U3m}.  Thus
the determination of the mixing angles yields at present a maximum
allowed CP violation
\begin{equation}
  \label{eq:jmax}
  J_\text{CP}^\text{max} = 0.0329 \pm 0.0007 \, (^{+0.0021}_{-0.0024})
\end{equation}
at $1\sigma$ ($3\sigma$) for both orderings.  The preference of the
present data for non-zero $\delta_\text{CP}$ implies a best fit value
$J_\text{CP}^\text{best} = -0.033$, which is favored over CP
conservation with $\Delta\chi^2 = 1.7$.  These numbers can be compared
with the size of the Jarlskog invariant in the quark sector, which is
determined to be $J_\text{CP}^\text{quarks} = (3.04^{+0.21}_{-0.20})
\times 10^{-5}$~\cite{PDG}.

\begin{figure}\centering
\includegraphics[width=0.9\textwidth]{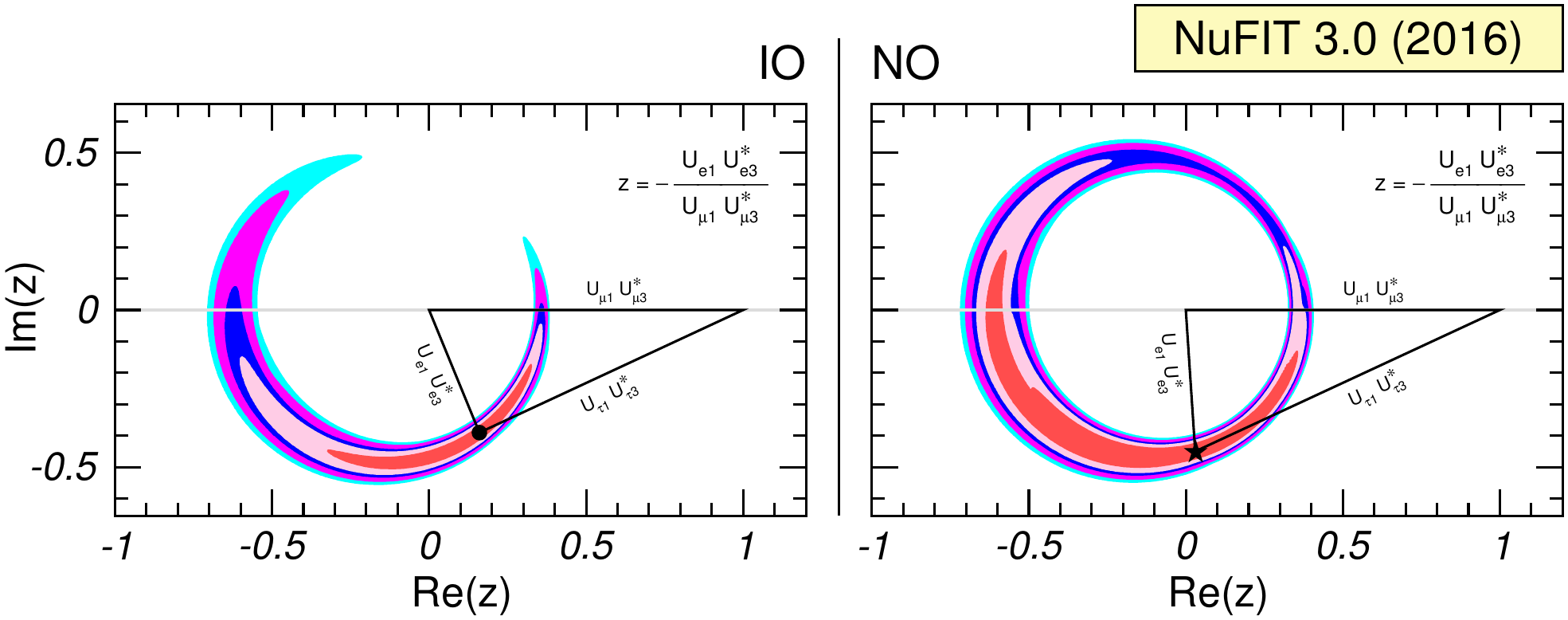}
  \caption{Leptonic unitarity triangle for the first and third columns
    of the mixing matrix.  After scaling and rotating the triangle so
    that two of its vertices always coincide with $(0,0)$ and $(1,0)$
    we plot the $1\sigma$, 90\%, $2\sigma$, 99\%, $3\sigma$ CL (2~dof)
    allowed regions of the third vertex. Note that in the construction
    of the triangle the unitarity of the $U$ matrix is always
    explicitly imposed. The regions for both orderings are defined
    with respect to the common global minimum which is in NO.}
  \label{fig:region-viola}
\end{figure}

In Fig.~\ref{fig:region-viola} we recast the allowed regions for the
leptonic mixing matrix in terms of one leptonic unitarity
triangle. Since in the analysis $U$ is unitary by construction, any
given pair of rows or columns can be used to define a triangle in the
complex plane. In the figure we show the triangle corresponding to the
unitarity conditions on the first and third columns which is the
equivalent to the one usually shown for the quark sector.  In this
figure the absence of CP violation implies a flat triangle,
\textit{i.e.}, $\Im(z) = 0$. As can be seen, for NO the horizontal
axis crosses the $1\sigma$ allowed region, which for 2~dof corresponds
to $\Delta\chi^2 \leq 2.3$. This is consistent with the present
preference for CP violation, $\chi^2(J_\text{CP} = 0) -
\chi^2(J_\text{CP}~\text{free}) = 1.7$ mentioned above. We will
comment on the statistical interpretation of this number in
Sec.~\ref{sec:MC}.

\section{Issues in present analysis}
\label{sec:issues}

The $3\nu$ fit results in the previous section provide a statistically
satisfactory description of all the neutrino oscillation data
considered.  There are however some issues in the determination of
some of the parameters which, although not of statistical significance
at present, deserve some attention.

\subsection{Status of $\Dmq_{21}$ in solar experiments versus KamLAND}
\label{subsec:dm12}

\begin{figure}\centering
\includegraphics[width=0.9\textwidth]{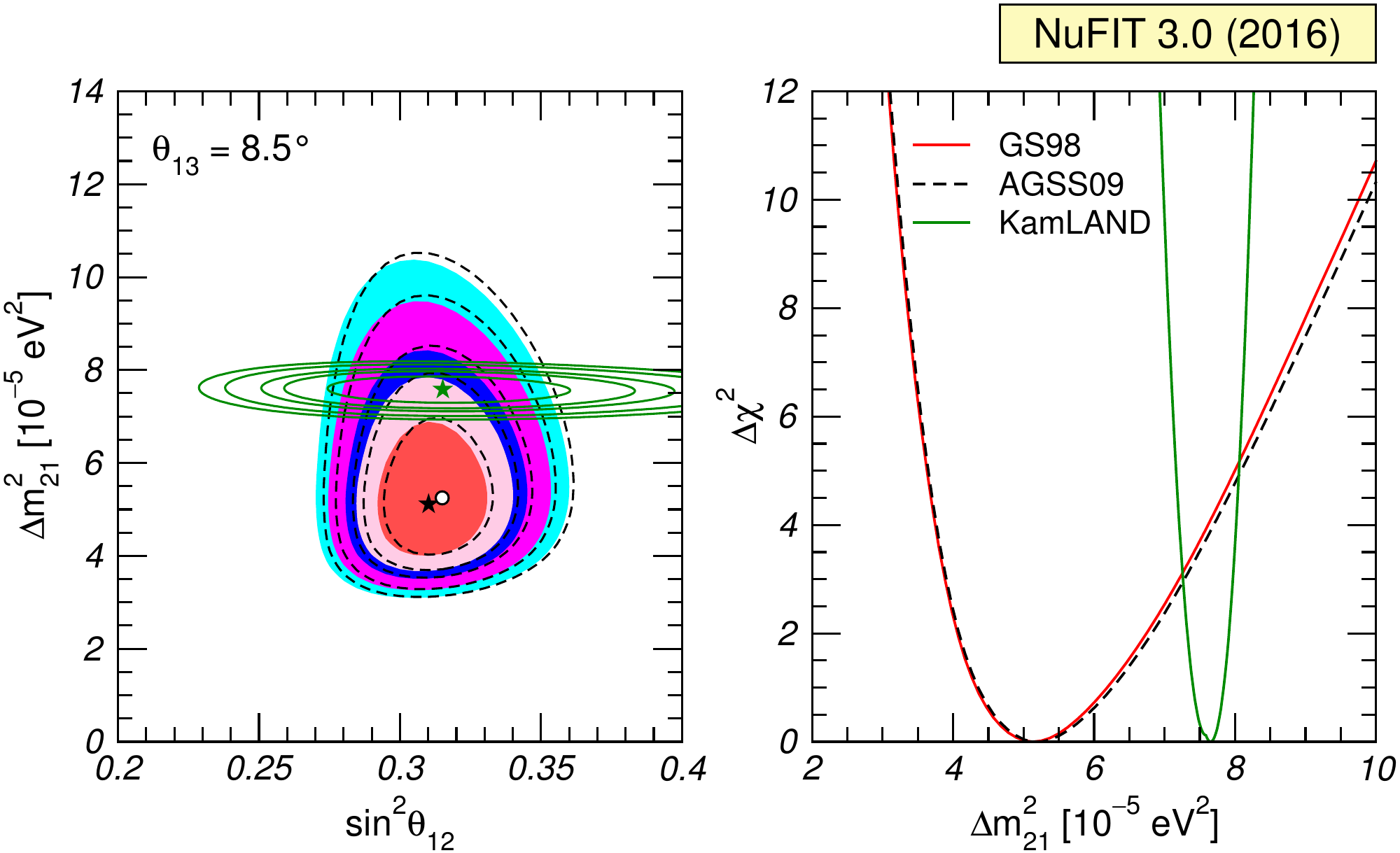}
  \caption{Left: Allowed parameter regions (at $1\sigma$, 90\%,
    $2\sigma$, 99\% and $3\sigma$ CL for 2~dof) from the combined
    analysis of solar data for GS98 model (full regions with best fit
    marked by black star) and AGSS09 model (dashed void contours with
    best fit marked by a white dot), and for the analysis of KamLAND
    data (solid green contours with best fit marked by a green star)
    for fixed $\theta_{13}=8.5^\circ$.  Right: $\Delta\chi^2$
    dependence on $\Dmq_{21}$ for the same three analyses after
    marginalizing over $\theta_{12}$.}
  \label{fig:sun-tension}
\end{figure}

The analyses of the solar experiments and of KamLAND give the dominant
contribution to the determination of $\Dmq_{21}$ and $\theta_{12}$.
It has been a result of global analyses for several years already,
that the value of $\Dmq_{21}$ preferred by KamLAND is somewhat higher
than the one from solar experiments. This tension arises from a
combination of two effects which have not changed significantly over
the last lustrum: a) the well-known fact that none of the $^8$B
measurements performed by SNO, SK and Borexino shows any evidence of
the low energy spectrum turn-up expected in the standard
LMA-MSW~\cite{Wolfenstein:1977ue, Mikheev:1986gs} solution for the
value of $\Dmq_{21}$ favored by KamLAND; b) the observation of a
non-vanishing day-night asymmetry in SK, whose size is larger than the
one predicted for the $\Dmq_{21}$ value indicated of KamLAND (for
which Earth matter effects are very small).  In
Ref.~\cite{Gonzalez-Garcia:2014bfa} we discussed the differences in
the physics entering in the analyses of solar and KamLAND data which
are relevant to this tension, and to which we refer the reader for
details. Here for sake of completeness we show in
Fig.~\ref{fig:sun-tension} the quantification of this tension in our
present global analysis. As seen in the figure, the best fit value of
$\Dmq_{21}$ of KamLAND lays at the boundary of the $2\sigma$ allowed
range of the solar neutrino analysis.

Also for illustration of the independence of these results with
respect to the solar modeling, the solar neutrino regions are shown
for two latest versions of the Standard Solar Model, namely the GS98
and the AGSS09 models~\cite{Bergstrom:2016cbh} obtained with two
different determinations of the solar
abundances~\cite{Vinyoles:2016djt}.

\subsection{$\Dmq_{3\ell}$ determination in LBL accelerator experiments
  versus reactors}
\label{subsec:dm32}

\begin{figure}\centering
\includegraphics[width=0.9\textwidth]{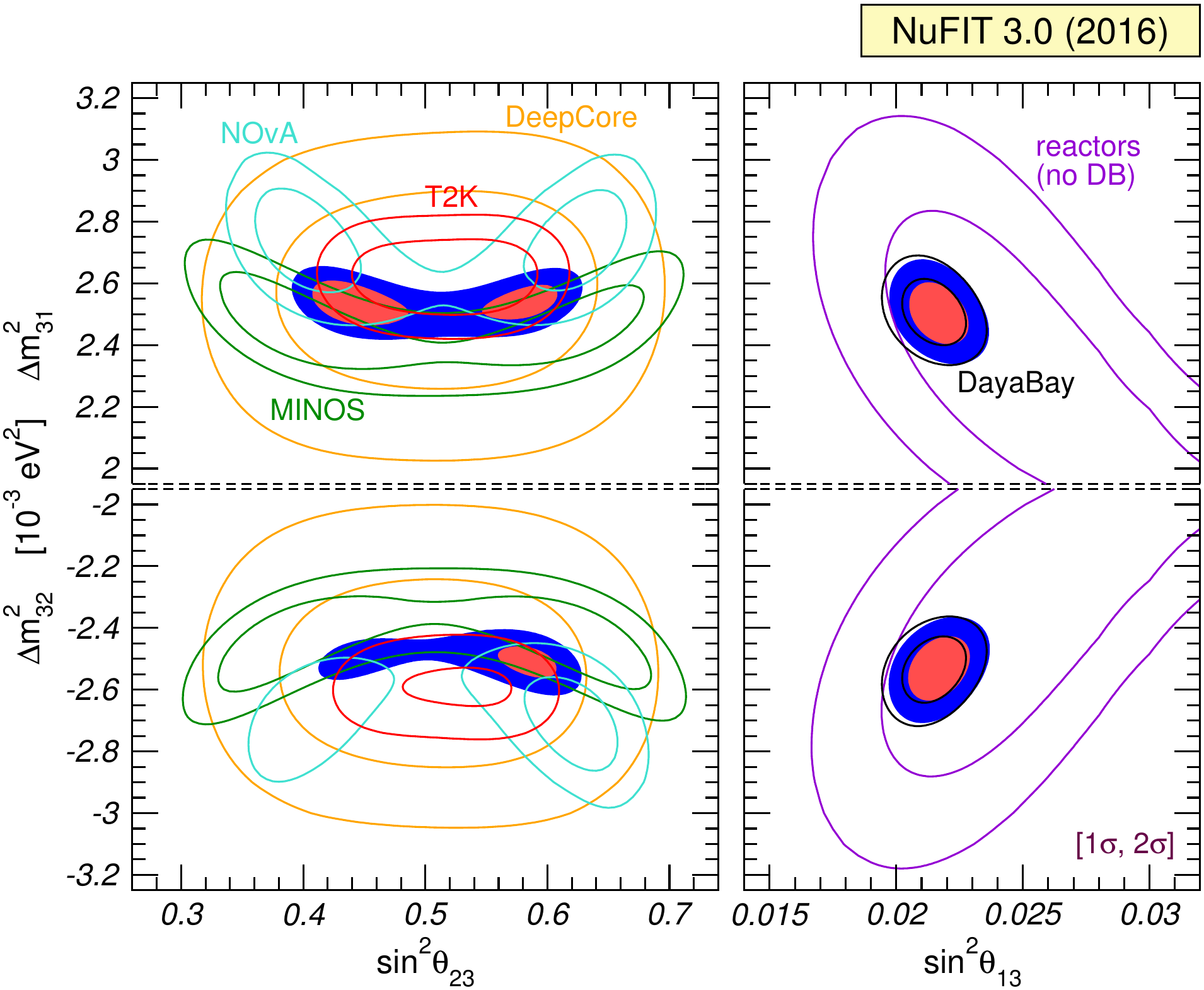}
  \caption{Determination of $\Dmq_{3\ell}$ at $1\sigma$ and $2\sigma$
    (2~dof), where $\ell=1$ for NO (upper panels) and $\ell=2$ for IO
    (lower panels). The left panels show regions in the $(\theta_{23},
    \Dmq_{3\ell})$ plane using both appearance and disappearance data
    from MINOS (green line), T2K (red lines), NO$\nu$A (light blue
    lines), as well as IceCube/DeepCore (orange lines) and the
    combination of them (colored regions). In these panels the
    constraint on $\theta_{13}$ from the global fit (which is
    dominated by the reactor data) is imposed as a Gaussian bias.  The
    right panels show regions in the $(\theta_{13}, \Dmq_{3\ell})$
    plane using only Daya-Bay (black lines), reactor data without
    Daya-Bay (violet lines), and their combination (colored
    regions). In all panels solar and KamLAND data are included to
    constrain $\Dmq_{21}$ and $\theta_{12}$. Contours are defined with
    respect to the global minimum of the two orderings.}
  \label{fig:region-sample}
\end{figure}

Figure~\ref{fig:region-sample} illustrates the contribution to the
present determination of $\Dmq_{3\ell}$ from the different data sets.
In the left panels we focus on the determination from long baseline
experiments, which is mainly from $\nu_\mu$ disappearance data. We
plot the $1\sigma$ and $2\sigma$ allowed regions (2~dof) in the
dominant parameters $\Dmq_{3\ell}$ and $\theta_{23}$. As seen in the
figure, although the agreement between the different experiments is
reasonable, some ``tension'' starts to appear in the determination of
both parameters among the LBL accelerator experiments.  In particular
we see that the recent results from NO$\nu$A, unlike those from T2K,
favor a non-maximal value of $\theta_{23}$. It is important to notice
that in the context of $3\nu$ mixing the relevant oscillation
probabilities for the LBL accelerator experiments also depend on
$\theta_{13}$ (and on the $\theta_{12}$ and $\Dmq_{21}$ parameters
which are independently well constrained by solar and KamLAND data).
To construct the regions plotted in the left panels of
Fig.~\ref{fig:region-sample}, we adopt the procedure currently
followed by the LBL accelerator experiments: we marginalize with
respect to $\theta_{13}$, taking into account the information from
reactor data by adding a Gaussian penalty term to the corresponding
$\chi^2_\text{LBL}$. This is not the same as making a combined
analysis of LBL and reactor data as we will quantify in
Sec.~\ref{subsec:t23ordcp}.

Concerning $\nu_e$ disappearance data, the total rates observed in
reactor experiments at different baselines can provide an independent
determination of $\Dmq_{3\ell}$~\cite{Bezerra:2012at,
  GonzalezGarcia:2012sz}.  On top of this, the observation of the
energy-dependent oscillation effect due to $\theta_{13}$ now allows to
further strengthen such measurement.  In the right panels of
Fig.~\ref{fig:region-sample} we show therefore the allowed regions in
the $(\theta_{13}, \Dmq_{3\ell})$ plane based on global data on
$\nu_e$ disappearance. The violet contours are obtained from all the
medium-baselines reactor experiments with the exception of Daya-Bay;
these regions emerge from the baseline effect mentioned above plus
spectral information from Double-Chooz.\footnote{Recently, RENO has
  presented a spectral analysis based on an exposure of 500
  days~\cite{Seo:2016uom}. Here we prefer to include from RENO only
  the total rate measurement, based on the larger exposure of 800
  days~\cite{reno:nu2014}.}  The black contours are based on the
energy spectrum in Daya-Bay, whereas the colored regions show the
combination.

\begin{figure}\centering
\includegraphics[width=0.8\textwidth]{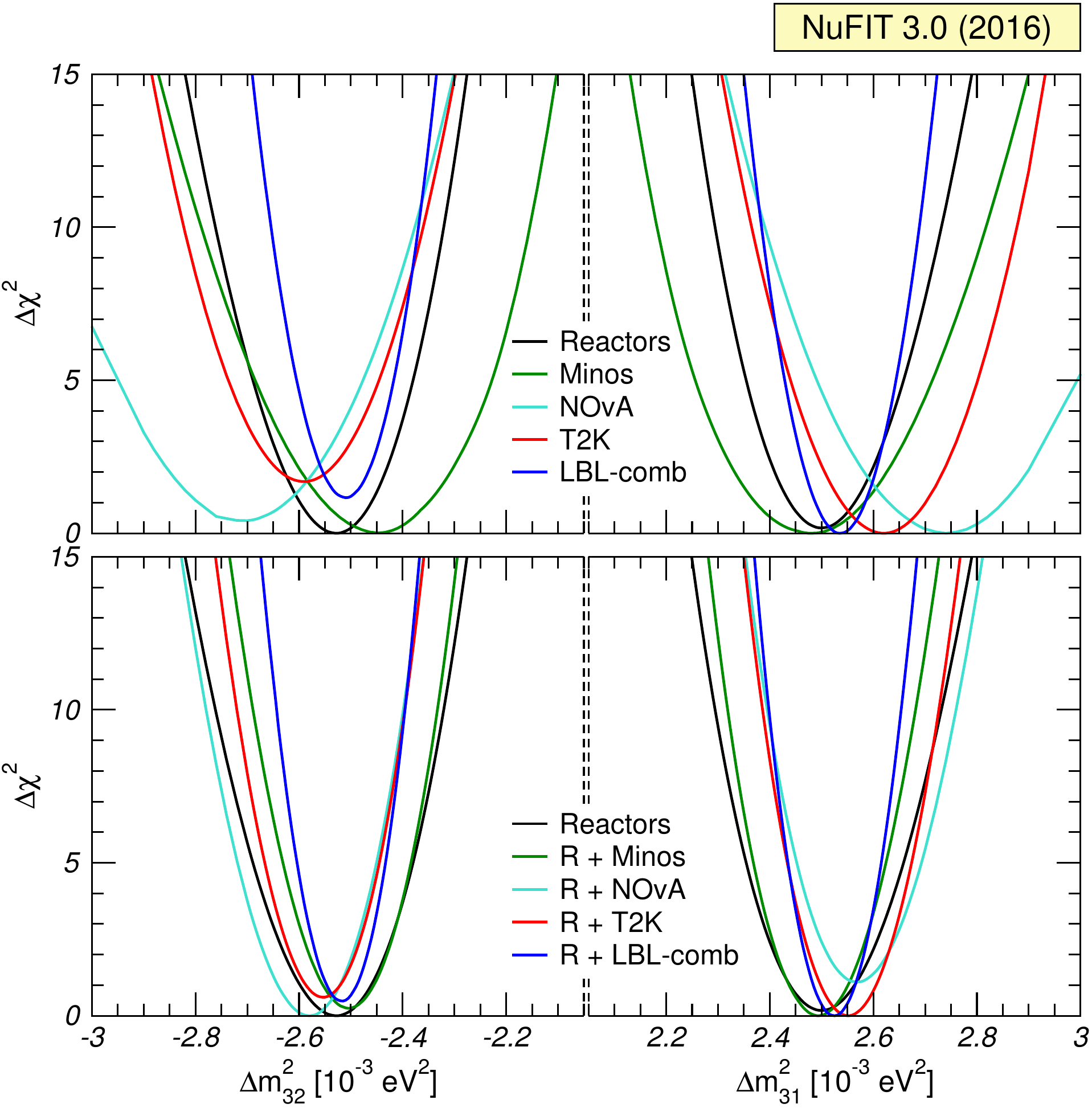}
\caption{$\Dmq_{3\ell}$ determination from LBL accelerator
  experiments, reactor experiments and their combination. Left (right)
  panels are for IO (NO). The upper panels show the 1-dim
  $\Delta\chi^2$ from LBL accelerator experiments after constraining
  \emph{only} $\theta_{13}$ from reactor experiments (this is,
  marginalizing Eq.~\eqref{eq:lblt13r} with respect to $\theta_{23}$
  and $\delta_\text{CP}$). For each experiment $\Delta\chi^2$ is
  defined with respect to the global minimum of the two orderings.
  The lower panels show the corresponding determination when the full
  information of LBL and reactor experiments is used in the
  combination (this is, marginalizing Eq.~\eqref{eq:lblreac} with
  respect to $\theta_{23}$ and $\delta_\text{CP}$).}
 \label{fig:chisq-dma}
\end{figure}

By comparing the left and right panels of Fig.~\ref{fig:region-sample}
we observe that the combined $\nu_\mu$ and $\nu_e$ disappearance
experiments provide a consistent determination of $|\Dmq_{3\ell}|$
with similar precision.  However when comparing the region for each
LBL experiment with that of the reactor experiments we find some
dispersion in the best fit values and allowed ranges.
This is more clearly illustrated in the upper panels of
Fig.~\ref{fig:chisq-dma}, where we plot the one dimensional projection
of the regions in Fig.~\ref{fig:region-sample} as a function of
$\Dmq_{3\ell}$ after marginalization over $\theta_{23}$ for each of
the LBL experiments and for their combination, together with that from
reactor data after marginalization over $\theta_{13}$.  The
projections are shown for NO(right) and IO(left). Let us stress that
the curves corresponding to LBL experiments in the upper panels of
Fig.~\ref{fig:chisq-dma} (as well as those in the upper panels of
Figs.~\ref{fig:chisq-t23} and~\ref{fig:chisq-dcp}) have been obtained
by a partial combination of the information on the shown parameter
($\Dmq_{3\ell}$ or $\theta_{23}$ or $\delta_\text{CP}$) from LBL with
that of $\theta_{13}$ from reactors, because in these plots only the
$\theta_{13}$ constraint from reactors is imposed while the dependence
on $\Dmq_{3\ell}$ is neglected. This corresponds to the 1-dim
projections of the function:
\begin{multline}
  \Delta\chi^2_\text{LBL+$\theta_{13}^\text{REA}$}
  (\theta_{23}, \delta_\text{CP}, \Dmq_{3\ell})
  \\
  = \min_{\theta_{13}} \Big[
    \chi^2_\text{LBL}(\theta_{13}, \theta_{23}, \delta_\text{CP}, \Dmq_{3\ell})
    + \min_{\Dmq_{3\ell}}\chi^2_\text{REA}(\theta_{13}, \Dmq_{3\ell}) \Big]
  - \chi^2_\text{min} \,.
  \label{eq:lblt13r}
\end{multline}

However, since reactor data also depends on $\Dmq_{3\ell}$ the full
combination of reactor and LBL results implies that one must add
consistently the $\chi^2$ functions of the LBL experiment with that of
reactors evaluated the same value of $\Dmq_{3\ell}$, this is
\begin{multline}
  \Delta\chi^2_\text{LBL+REA}
  (\theta_{23}, \delta_\text{CP}, \Dmq_{3\ell})
  \\
  = \min_{\theta_{13}} \Big[
    \chi^2_\text{LBL}(\theta_{13}, \theta_{23}, \delta_\text{CP}, \Dmq_{3\ell})
    + \chi^2_\text{REA}(\theta_{13}, \Dmq_{3\ell}) \Big]
  - \chi^2_\text{min} \,.
  \label{eq:lblreac}
\end{multline}
We discuss next the effect of combining consistently the information
from LBL and reactor experiments in the present determination of
$\theta_{23}$, $\delta_\text{CP}$ and the ordering.

\subsubsection{Impact on the determination of $\theta_{23}$, mass ordering, and
  $\delta_\text{CP}$}
\label{subsec:t23ordcp}

\begin{figure}\centering
  \includegraphics[width=0.8\textwidth]{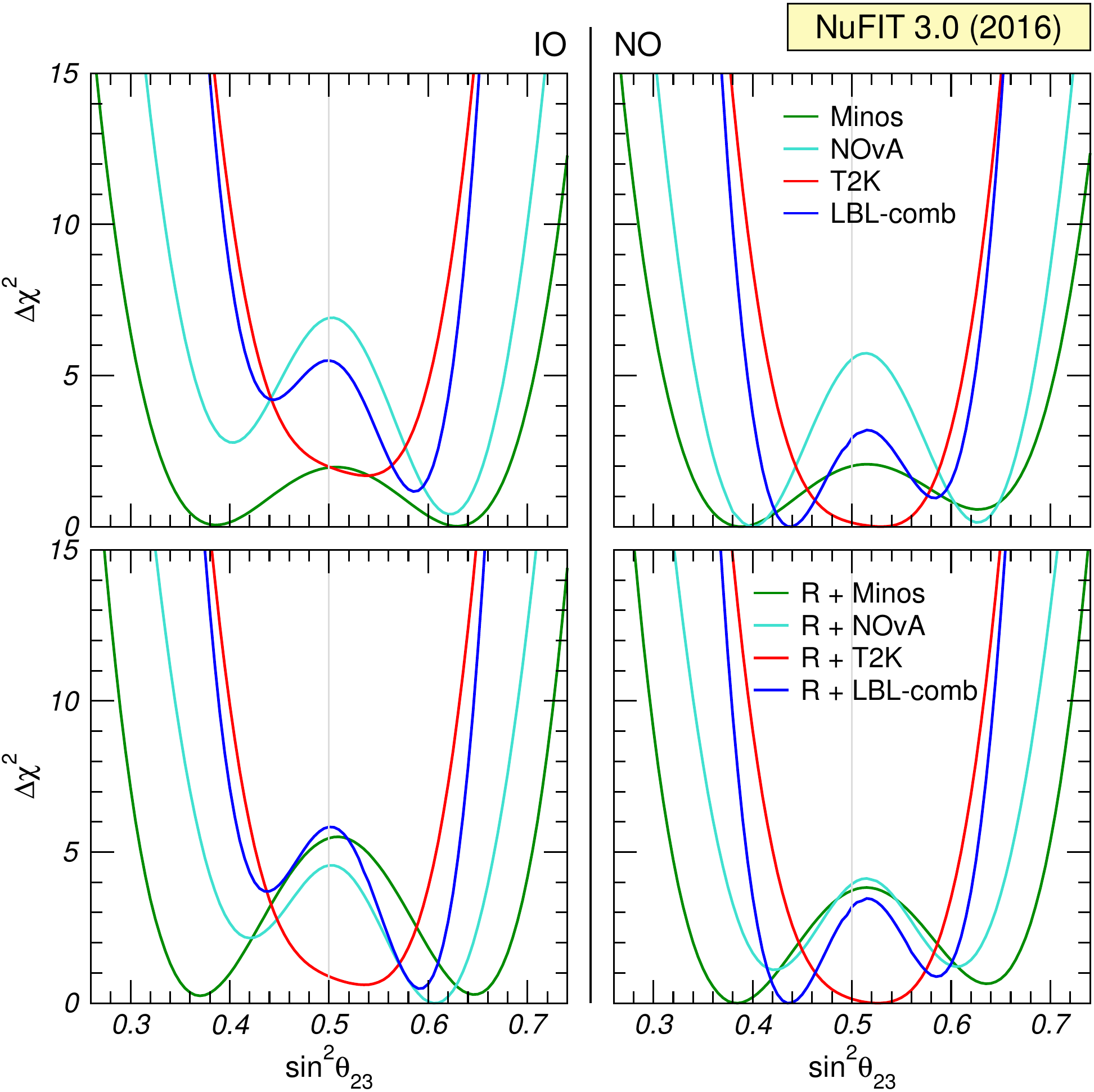}
  \caption{$\theta_{23}$ determination from LBL, reactor and their
    combination. Left (right) panels are for IO (NO). The upper panels
    show the 1-dim $\Delta\chi^2$ from LBL experiments after
    constraining \emph{only} $\theta_{13}$ from reactor experiments
    (this is, marginalizing Eq.~\eqref{eq:lblt13r} with respect to
    $\Dmq_{3\ell}$ and $\delta_\text{CP}$). For each experiment
    $\Delta\chi^2$ is defined with respect to the global minimum of
    the two orderings.  The lower panels show the corresponding
    determination when the full information of LBL accelerator and
    reactor experiments is used in the combination (this is,
    marginalizing Eq.~\eqref{eq:lblreac} with respect to
    $\Dmq_{3\ell}$ and $\delta_\text{CP}$).}
  \label{fig:chisq-t23}
\end{figure}

\begin{figure}\centering
  \includegraphics[width=0.8\textwidth]{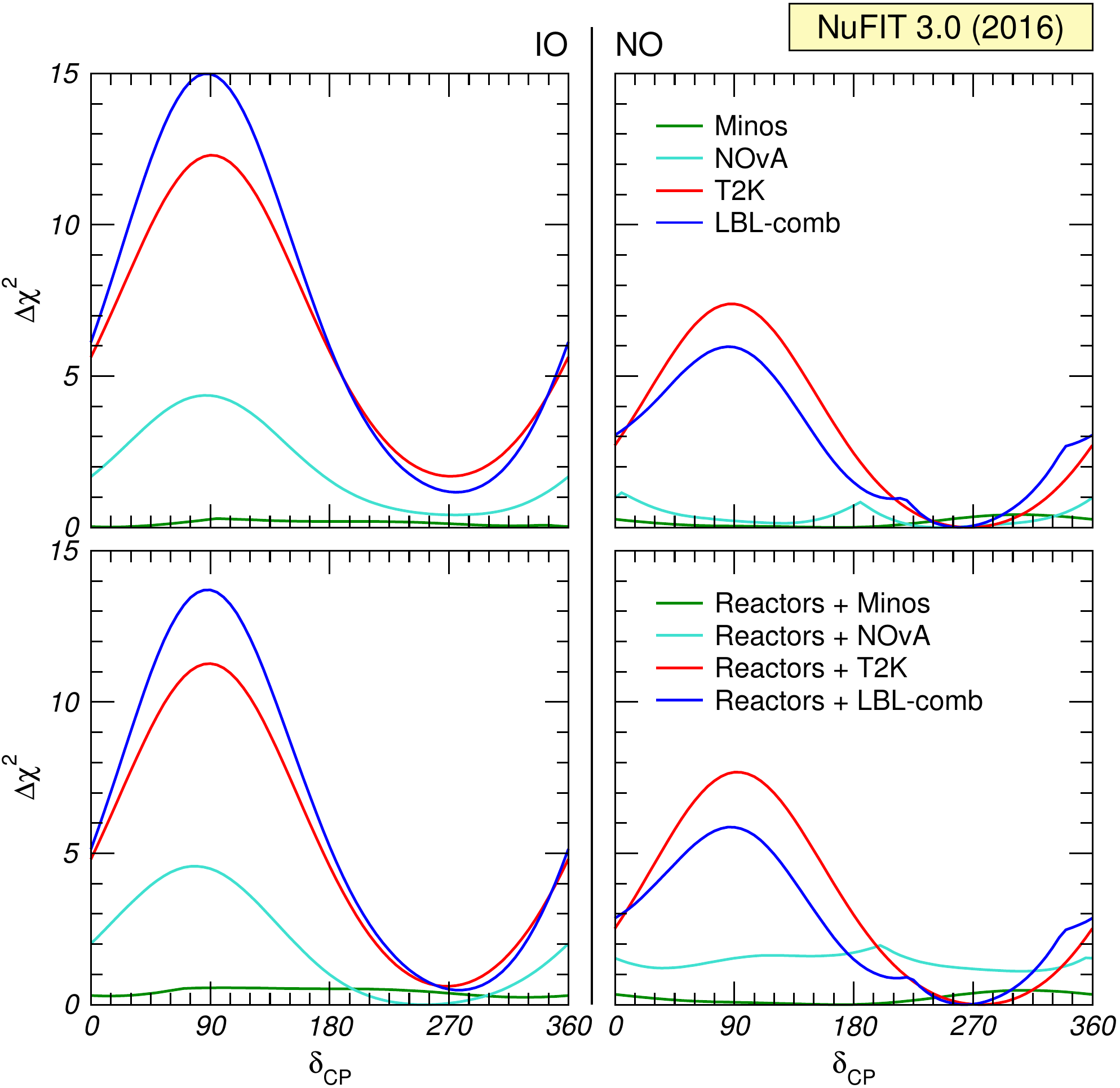}
  \caption{$\delta_\text{CP}$ determination from LBL, reactor and
    their combination. Left (right) panels are for IO (NO). The upper
    panels show the 1-dim $\Delta\chi^2$ from LBL experiments after
    constraining \emph{only} $\theta_{13}$ from reactor experiments
    (this is, marginalizing Eq.~\eqref{eq:lblt13r} with respect to
    $\Dmq_{3\ell}$ and $\theta_{23}$). For each experiment
    $\Delta\chi^2$ is defined with respect to the global minimum of
    the two orderings.  The lower panels show the corresponding
    determination when the full information of LBL accelerator and
    reactor experiments is used in the combination (this is,
    marginalizing Eq.~\eqref{eq:lblreac} with respect to
    $\Dmq_{3\ell}$ and $\theta_{23}$).}
  \label{fig:chisq-dcp}
\end{figure}

We plot in the lower panels of
Figs.~\ref{fig:chisq-dma}--\ref{fig:chisq-dcp} the one dimensional
projections of $\Delta\chi^2_\text{LBL+REA}$ for each of the
parameters $\theta_{23}$, $\delta_\text{CP}$, $\Dmq_{3\ell}$
(marginalized with respect to the two undisplayed parameters) for the
consistent LBL+REA combinations with both the information on
$\theta_{13}$ and $\Dmq_{3\ell}$ from reactors included,
Eq.~\eqref{eq:lblreac}.  As mentioned before, the curves in the upper
panels for these figures show the corresponding 1-dimensional
projections for the partial combination, in which only the
$\theta_{13}$ constraint from reactors is used,
Eq.~\eqref{eq:lblt13r}.  For each experiment the curves in these
figures are defined with respect to the global minimum of the two
orderings, so the relative height of the minimum in one ordering vs
the other gives a measure of the ordering favored by each of the
experiments.

Comparing the upper and lower panels in Figs.~\ref{fig:chisq-dma},
\ref{fig:chisq-t23} and~\ref{fig:chisq-dcp} one sees how the
contribution to the determination of the mass ordering, the octant and
non-maximality of $\theta_{23}$, and the presence of leptonic CP
violation of each LBL experiment in the full LBL+REA combination
(Eq.~\ref{eq:lblreac}) can differ from those derived from the LBL
results imposing only the $\theta_{13}$ constraint from reactors
(Eq.~\ref{eq:lblt13r}). This is due to the additional information on
$\Dmq_{3\ell}$ from reactors, which is missing in this last case.  In
particular:
\begin{itemize}
\item When only combining the results of the accelerator LBL
  experiments with the reactor bound of $\theta_{13}$, both NO$\nu$A
  and T2K favor NO by
  $\chi^2_\text{LBL+$\theta_{13}^\text{REA}$}(\text{IO})
  -\chi^2_\text{LBL+$\theta_{13}^\text{REA}$}(\text{NO}) \simeq 0.4$
  ($1.7$) for $\text{LBL} = \text{NO$\nu$A}$ (T2K). This is in
  agreement with the analyses shown by the collaborations for example
  in Refs.~\cite{nova:nu2016, t2k:ichep2016}.  However, when
  consistently combining with the reactor data, we find that the
  preference for NO by T2K+REA is reduced, and NO$\nu$A+REA actually
  favors IO. This is due to the slightly lower value of
  $|\Dmq_{3\ell}|$ favored by the reactor data, in particular in
  comparison with NO$\nu$A for both orderings, and also with T2K for
  NO.  Altogether we find that for the full combination of LBL
  accelerator experiments with reactors the ``hint'' towards NO is
  below $1\sigma$.

\item Figure~\ref{fig:chisq-t23} illustrates how both NO$\nu$A and
  MINOS favor non-maximal $\theta_{23}$.  From this figure we see that
  while the significance of non-maximality in NO$\nu$A seems more
  evident than in MINOS when only the information of $\theta_{13}$ is
  included (upper panels), the opposite holds for the full combination
  with the reactor data (lower panels). In particular,
  \begin{equation}
    \begin{aligned}
      \chi^2_\text{LBL+$\theta_{13}^\text{REA}$}(\theta_{23}=45^\circ,\text{NO})
      - \min_{\theta_{23}}
        \chi^2_\text{LBL+$\theta_{13}^\text{REA}$}(\theta_{23},\text{NO})
      &= 5.5~(2.0) \,,
      \\
      \chi^2_\text{LBL+$\theta_{13}^\text{REA}$}(\theta_{23}=45^\circ,\text{IO})
      - \min_{\theta_{23}}
        \chi^2_\text{LBL+$\theta_{13}^\text{REA}$}(\theta_{23},\text{IO})
      &= 6.5~(1.9) \,,
      \\
      \chi^2_\text{LBL+REA}(\theta_{23}=45^\circ,\text{NO})
      - \min_{\theta_{23}}
        \chi^2_\text{LBL+REA}(\theta_{23},\text{NO})
      &= 2.8~(3.7) \,,
      \\
      \chi^2_\text{LBL+REA}(\theta_{23}=45^\circ,\text{IO})
      - \min_{\theta_{23}}
        \chi^2_\text{LBL+REA}(\theta_{23},\text{IO})
      &= 4.6~(5.2) \,,
    \end{aligned}
  \end{equation}
  for LBL = NO$\nu$A (MINOS).  On the other hand T2K results are
  compatible with $\theta_{23} = 45^\circ$ for any
  ordering. Altogether we find that for NO the full combination of LBL
  accelerator experiments and reactors disfavor maximal $\theta_{23}$
  mixing by $\Delta\chi^2 = 3.2$.

\item Regarding the octant of $\theta_{23}$, for IO all LBL
  accelerator experiments are better described with $\theta_{23} >
  45^\circ$, adding up to a $\sim 1.8\sigma$ preference for that
  octant. Conversely, for NO $\theta_{23} < 45^\circ$ is favored at
  $\sim 1\sigma$.

\item From Fig.~\ref{fig:chisq-dcp} we see that the ``hint'' for a CP
  phase around $270^\circ$ is mostly driven by T2K data, with some
  extra contribution from NO$\nu$A in the case of IO.  Within the
  present precision the favored ranges of $\delta_\text{CP}$ in each
  ordering by the combination of LBL accelerator experiments are
  pretty independent on the inclusion of the $\Dmq_{3\ell}$
  information from reactors.
\end{itemize}

\subsection{Analysis of Super-Kamiokande atmospheric data}
\label{subsec:SK}

\begin{figure}\centering
  \includegraphics[width=0.9\textwidth]{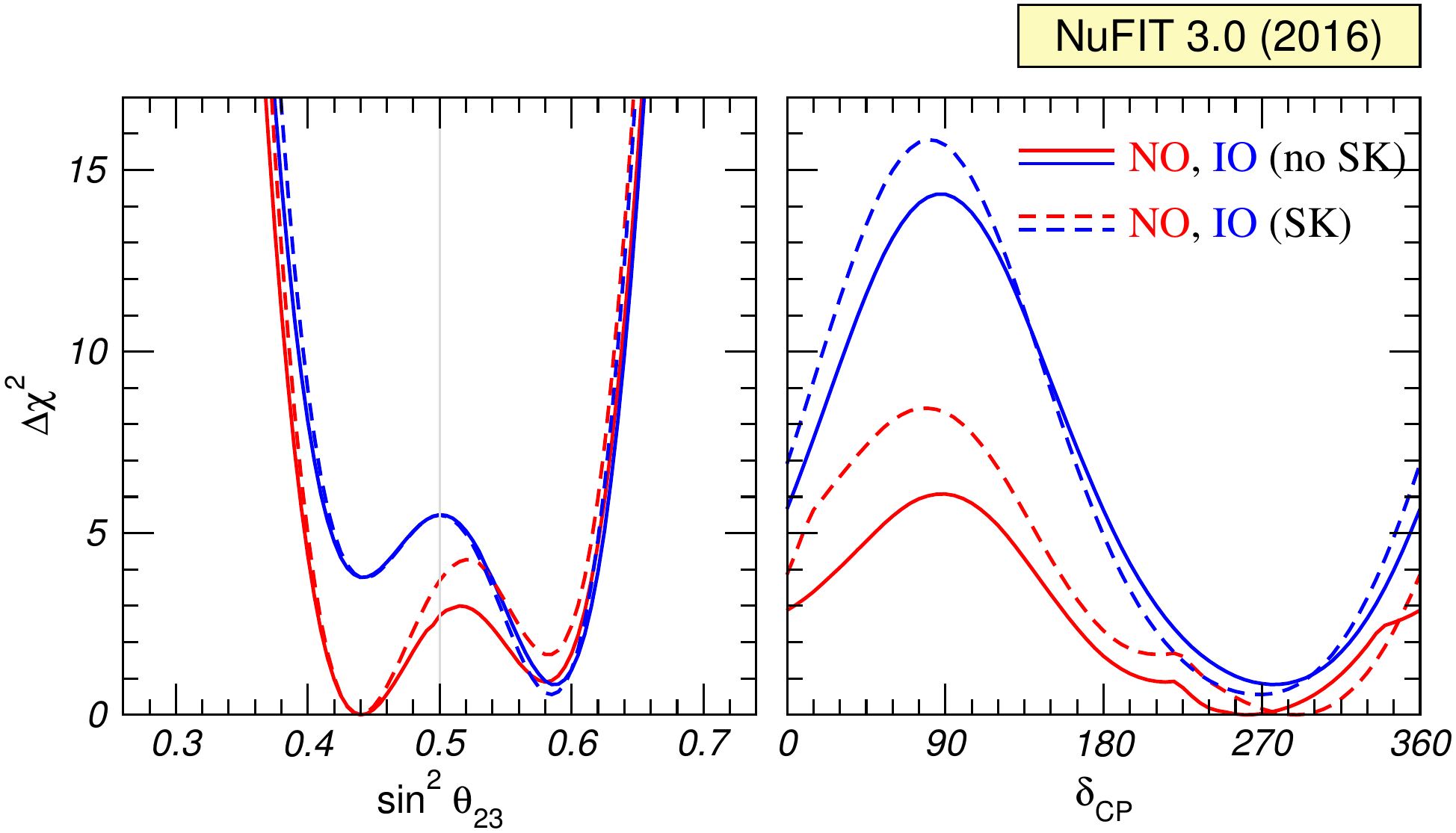}
  \caption{Impact of our re-analysis of SK atmospheric neutrino
    data~\cite{Wendell:2014dka} (70 bins in energy and zenith angle)
    on the determination of $\sin^2\theta_{23}$, $\delta_\text{CP}$,
    and the mass ordering. The impact on all other parameters is
    negligible.}
  \label{fig:chisq-atmos}
\end{figure}

In all the results discussed so far we have not included information
from Super-Kamiokande atmospheric data. The reason is that our
oscillation analysis cannot reproduce that of the collaboration
presented in their talks in the last two years (see for example
Ref.~\cite{skatm:nufact2016} for their latest unpublished results).

\begin{table}\centering
  \begin{footnotesize}
    \begin{tabular}{l|cc|cc|c}
      \hline\hline
      & \multicolumn{2}{c|}{Normal Ordering (best fit)}
      & \multicolumn{2}{c|}{Inverted Ordering ($\Delta\chi^2=0.56$)}
      & Any Ordering
      \\
      \hline
      & bfp $\pm 1\sigma$ & $3\sigma$ range
      & bfp $\pm 1\sigma$ & $3\sigma$ range
      & $3\sigma$ range
      \\
      \hline
      \rule{0pt}{4mm}\ignorespaces
      $\sin^2\theta_{23}$
      & $0.440_{-0.019}^{+0.024}$ & $0.388 \to 0.630$
      & $0.584_{-0.022}^{+0.019}$ & $0.398 \to 0.634$
      & $0.388 \to 0.632$
      \\[1mm]
      $\theta_{23}/^\circ$
      & $41.5_{-1.1}^{+1.4}$ & $38.6 \to 52.5$
      & $49.9_{-1.3}^{+1.1}$ & $39.1 \to 52.8$
      & $38.6 \to 52.7$
      \\[3mm]
      $\delta_\text{CP}/^\circ$
      & $289_{-51}^{+38}$ & $\hphantom{00}0 \to 360$
      & $269_{-45}^{+40}$ & $146 \to 377$
      & $\hphantom{00}0 \to 360$
      \\[1mm]
      \hline\hline
    \end{tabular}
  \end{footnotesize}
  \caption{Three-flavor oscillation parameters from our fit to global
    data, including also our re-analysis of SK1--4 (4581 days)
    atmospheric data.  The numbers in the 1st (2nd) column are
    obtained assuming NO (IO), \textit{i.e.}, relative to the
    respective local minimum, whereas in the 3rd column we minimize
    also with respect to the ordering.  The omitted parameters are
    identical to Tab.~\ref{tab:bfranges}.}
  \label{tab:skranges}
\end{table}

Already since SK2 the Super-Kamiokande collaboration has been
presenting its experimental results in terms of a growing number of
data samples.  The rates for some of those samples cannot be predicted
(and therefore included in a statistical analysis) without a detailed
simulation of the detector, which can only be made by the experimental
collaboration itself. Our analysis of Super-Kamiokande data has been
always based on the ``classical'' set of samples for which our
simulations were reliable enough: sub-GeV and multi-GeV $e$-like and
$\mu$-like fully contained events, as well as partially contained,
stopping and through-going muon data, each divided into 10 angular
bins for a total of 70 energy and zenith angle bins (details on our
simulation of the data samples and the statistical analysis are given
in the Appendix of Ref.~\cite{GonzalezGarcia:2007ib}).  Despite the
limitations, until recently our results represented the most
up-to-date analysis of the atmospheric neutrino data which could be
performed outside the collaboration, and we were able to reproduce
with reasonable precision the oscillation results of the full analysis
presented by SK~-- both for what concerns the determination of the
dominant parameters $\Dmq_{3\ell}$ and $\theta_{23}$, as well as their
rather marginal sensitivity to the subdominant $\nu_e$ appearance
effects driven by $\theta_{13}$ (and consequently to
$\delta_\text{CP}$ and the ordering). Thus we confidently included our
own implementation of the Super-Kamiokande $\chi^2$ in our global fit.

However, in the last two years Super-Kamiokande has developed a new
analysis method in which a set of neural network based selections are
introduced, some of them with the aim of constructing $\nu_e +
\bar\nu_e$ enriched samples which are then further classified into
$\nu_e$-like and $\bar\nu_e$-like subsamples, thus increasing the
sensitivity to subleading parameters such as the mass ordering and
$\delta_\text{CP}$~\cite{Wendell:2014dka, skatm:thesis}.  The
selection criteria are constructed to exploit the expected differences
in the number of charged pions and transverse momentum in the
interaction of $\nu_e$ versus $\bar\nu_e$.  With this new analysis
method Super-Kamiokande has been reporting in talks an increasing
sensitivity to the ordering and to $\delta_\text{CP}$: for example,
the preliminary results of the analysis of SK1--4 (including 2520 days
of SK4)~\cite{skatm:nufact2016} in combination with the reactor
constraint of $\theta_{13}$ show a preference for NO with a
$\Delta\chi^2(\text{IO}) = 4.3$ and variation of
$\chi^2(\delta_\text{CP})$ with the CP phase at the level of $\sim
1.7\sigma$.

Unfortunately, with publicly available information this analysis is
not reproducible outside the collaboration.  Conversely our
``traditional'' analysis based on their reproducible data samples
continues to show only marginal dependence on these effects. This is
illustrated in Fig.~\ref{fig:chisq-atmos} and Tab.~\ref{tab:skranges}
where we show the impact of inclusion of our last re-analysis of SK
atmospheric data using the above mentioned 70 bins in energy and
zenith angle.\footnote{We use the same data and statistical treatment
  as in our previous global fit
  NuFIT 2.0~\cite{Gonzalez-Garcia:2014bfa} as well as in versions 2.1
  and 2.2~\cite{nufit} which is based on 4581 days of data from
  SK1--4~\cite{Wendell:2014dka} (corresponding to 1775 days of SK4).}
We only show the impact on the determination of $\sin^2\theta_{23}$,
$\delta_\text{CP}$, and the mass ordering as the effect on all other
parameters is negligible. We observe that $\Delta\chi^2$ for maximal
mixing and the second $\theta_{23}$ octant receive an additional
contribution of about 1 unit in the case of NO, whereas the
$\theta_{23}$ result for IO is practically unchanged. Values of
$\delta_\text{CP} \simeq 90^\circ$ are slightly more disfavoured,
whereas there is basically no effect on the mass ordering
discrimination.

In summary, with the information at hand we are not able to reproduce
the elements driving the main dependence on the subdominant effects of
the official (though preliminary and unpublished) Super-Kamiokande
results, while the dominant parameters are currently well determined
by LBL experiments. For these reasons we have decided not to include
our re-analysis of Super-Kamiokande data in our preferred global fit
presented in the previous section.  Needless to say that when enough
quantitative information becomes available to allow a reliable
simulation of the subdominant $\nu_e$-driven effects, we will proceed
to include it in our global analysis.

\section{Monte Carlo evaluation of confidence levels  for $\theta_{23}$,
 $\delta_\text{CP}$ and ordering}
\label{sec:MC}

At present the three least known neutrino oscillation parameters are
the Dirac CP violating phase $\delta_\text{CP}$, the octant of
$\theta_{23}$ and the mass ordering (which in what follows we will
denote by ``O'').  In order to study the information from data on
these parameters one can use two $\Delta\chi^2$ test
statistics~\cite{Elevant:2015ska, Blennow:2014sja}:
\begin{align}
  \label{eq:testStatistics1}
  \Delta\chi^2 \left(\delta_\text{CP}, \text{O} \right)
  &= \min_{x_1}
  \chi^2\left(\delta_\text{CP}, \text{O}, x_1\right) -\chi^2_\text{min} \,,
  \\
  \label{eq:testStatistics2}
  \Delta \chi^2\left(\theta_{23}, \text{O}\right)
  &= \min_{x_2}
  \chi^2\left(\theta_{23}, \text{O}, x_2\right) -\chi^2_\text{min} \,,
\end{align}
where the minimization in the first equation is performed with respect
to all oscillation parameters except $\delta_\text{CP}$ and the
ordering ($x_1 = \lbrace \theta_{12}, \theta_{13}, \theta_{23},
\Dmq_{21}, |\Dmq_{3\ell}| \rbrace$), while in the second equation the
minimization is over all oscillation parameters except $\theta_{23}$
and the ordering ($x_2 = \lbrace \theta_{12}, \theta_{13},
\delta_\text{CP}, \Dmq_{21}, |\Dmq_{3\ell}| \rbrace$). Here
$\chi^2_\text{min}$ indicates the $\chi^2$ minimum with respect to all
oscillation parameters including the mass ordering.

We have plotted the values of these test statistics in the lower right
and central left panels in Fig.~\ref{fig:chisq-glob}. We can use them
not only for the determination of $\delta_\text{CP}$ and
$\theta_{23}$, respectively, but also of the mass ordering. For
instance, using Eq.~\eqref{eq:testStatistics1} we can determine a
confidence interval for $\delta_\text{CP}$ at a given CL for both
orderings. However, below a certain CL no interval will appear for the
less favored ordering. In this sense we can exclude that ordering at
the CL at which the corresponding interval for $\delta_\text{CP}$
disappears. Note that a similar prescription to test the mass ordering
can be built for any other parameter as well, \textit{e.g.}, for
$\theta_{23}$ using Eq.~\eqref{eq:testStatistics2}.\footnote{Let us
  mention that this method to determine the mass ordering is different
  from the one based on the test statistics $T$ discussed in
  Ref.~\cite{Blennow:2013oma}.}

\begin{figure}\centering
  \includegraphics[width=0.8\textwidth]{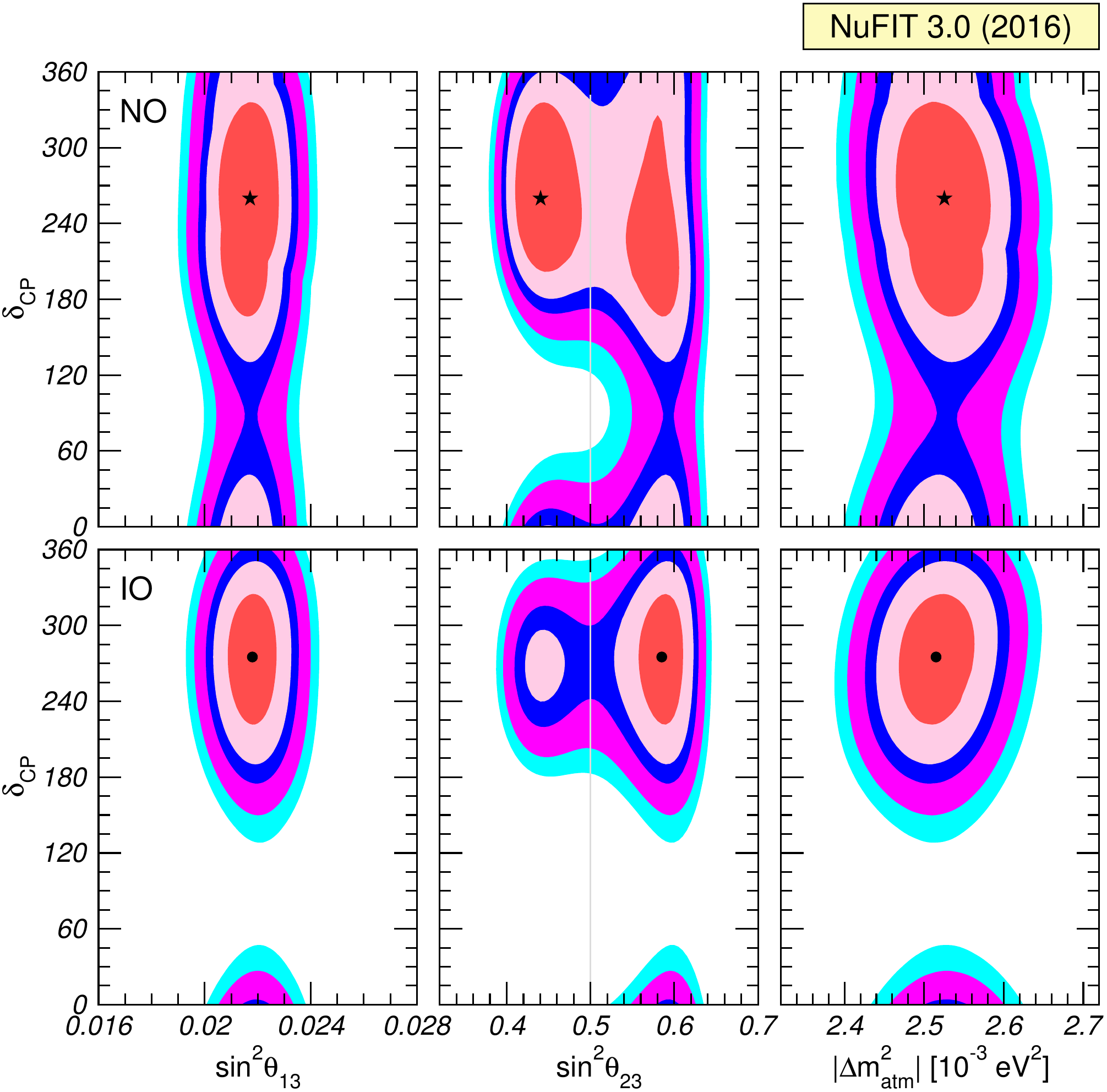}
  \caption{Allowed regions from the global data at $1\sigma$, 90\%,
    $2\sigma$, 99\% and $3\sigma$ CL (2~dof). We show projections onto
    different planes with $\delta_\text{CP}$ on the vertical axis
    after minimizing with respect to all undisplayed parameters. The
    lower (upper) panels correspond to IO (NO).  Contour regions are
    derived with respect to the global minimum which occurs for NO and
    is indicated by a star. The local minimum for IO is shown by a
    black dot.}
  \label{fig:region-hier}
\end{figure}

In Sec.~\ref{sec:global} we have presented confidence intervals
assuming that the test statistics follow a $\chi^2$-distribution with
1~dof, relying on Wilks theorem to hold~\cite{Wilks:1938dza} (this is
what we call the Gaussian limit). However, the test statistics in
Eqs.~\eqref{eq:testStatistics1} and~\eqref{eq:testStatistics2} are
expected not to follow Wilks' theorem because of several
reasons~\cite{Elevant:2015ska}:
\begin{itemize}
\item Sensitivity of current data to $\delta_\text{CP}$ is still
  limited, as can be seen in Fig.~\ref{fig:chisq-glob}: all values of
  $\delta_\text{CP}$ have $\Delta \chi^2 < 14$, and for NO not even
  $\Delta \chi^2 = 6$ is attained.

\item Regarding $\theta_{23}$, its precision is dominated by $\nu_\mu$
  disappearance experiments. Since the relevant survival probability
  depends dominantly on $\sin^2 2\theta_{23}$, there is both a
  physical boundary of their parameter space at $\theta_{23} =
  45^\circ$ (because $\sin 2\theta_{23}<1$), as well as a degeneracy
  related to the octant.

\item The mass ordering is a discrete parameter.

\item The dependence of the theoretical predictions on
  $\delta_\text{CP}$ is significantly non-linear, even more
  considering the periodic nature of this parameter. Furthermore,
  there are complicated correlations and degeneracies between
  $\delta_\text{CP}$, $\theta_{23}$, and the mass ordering (see
  Fig.~\ref{fig:region-hier} for illustration).
\end{itemize}
Therefore, one may expect deviations from the Gaussian limit of the
$\Delta\chi^2$ distributions, and confidence levels for these
parameters should be cross checked through a Monte Carlo simulation of
the relevant experiments. We consider in the following the combination
of the T2K, NO$\nu$A, MINOS and Daya-Bay experiments, which are most
relevant for the parameters we are interested in this section. For a
given point of assumed true values for the parameters we generate a
large number ($10^4$) of pseudo-data samples for each of the
experiments. For each pseudo-data sample we compute the two statistics
given in Eqs.~\eqref{eq:testStatistics1}
and~\eqref{eq:testStatistics2} to determine their distributions
numerically. In Ref.~\cite{Elevant:2015ska} it has been shown that the
distribution of test statistics for 2-dimensional parameter region
(such as for instance the middle panels of Fig.~\ref{fig:region-hier})
are more close to Gaussianity than 1-dimensional ones such as
Eqs.~\eqref{eq:testStatistics1}
and~\eqref{eq:testStatistics2}. Therefore we focus here on the
1-dimensional cases.

First, let us note that in order to keep calculation time manageable
one can fix all parameters which are known to be uncorrelated with the
three we are interested in (\textit{i.e.}, $\theta_{23}$,
$\delta_\text{CP}$, O).  This is certainly the case for $\Dmq_{21}$
and $\theta_{12}$ which are determined independently by solar and
KamLAND data. As for $\theta_{13}$, presently the most precise
information arises from reactor data whose results are insensitive to
$\delta_\text{CP}$ and $\theta_{23}$.  Consequently, marginalizing over
$\theta_{13}$ within reactor uncertainties or fixing it to the best
fit value gives a negligible difference in the simulations. Concerning
$|\Dmq_{3\ell}|$ we observe that there are no strong correlations or
degeneracies with $\delta_\text{CP}$ (see Fig.~\ref{fig:region-hier}),
and we assume that the distributions of the test statistics do not
significantly depend on the assumed true value. Therefore we consider
only the global best fit values for each ordering as true values for
$|\Dmq_{3\ell}|$ to generate pseudo-data. However, since the relevant
observables \emph{do} depend non-trivially on its value, it is
important to keep $|\Dmq_{3\ell}|$ as a free parameter in the fit and
to minimize the $\chi^2$ for each pseudo-data sample with respect to
it.  Hence, we approximate the test statistics in
Eqs.~\eqref{eq:testStatistics1} and~\eqref{eq:testStatistics2} by
using
\begin{align}
  \chi^2 \left(\delta_\text{CP}, \text{O}, x_1\right)
  &\equiv
  \min_{\theta_{23}, |\Dmq_{3\ell}|}
  \chi^2 \left(\theta_{23}, \delta_\text{CP}, \text{O}, |\Dmq_{3\ell}|\right)  \,,
  \\
  \chi^2 \left(\theta_{23}, \text{O}, x_2 \right)
  &\equiv
  \min_{\delta_\text{CP}, |\Dmq_{3\ell}|}
  \chi^2 \left(\theta_{23}, \delta_\text{CP}, \text{O}, |\Dmq_{3\ell}| \right) \,,
\end{align}
with the other oscillation parameters kept fixed at their best fit
points: $\Dmq_{21} = 7.5 \times 10^{-5}~\text{eV}^2$,
$\sin^2\theta_{12} = 0.31$, and $\sin^2\theta_{13} = 0.022$.

\subsection{$\delta_\text{CP}$ and the mass ordering}

\begin{figure}\centering
  \includegraphics[width=\textwidth]{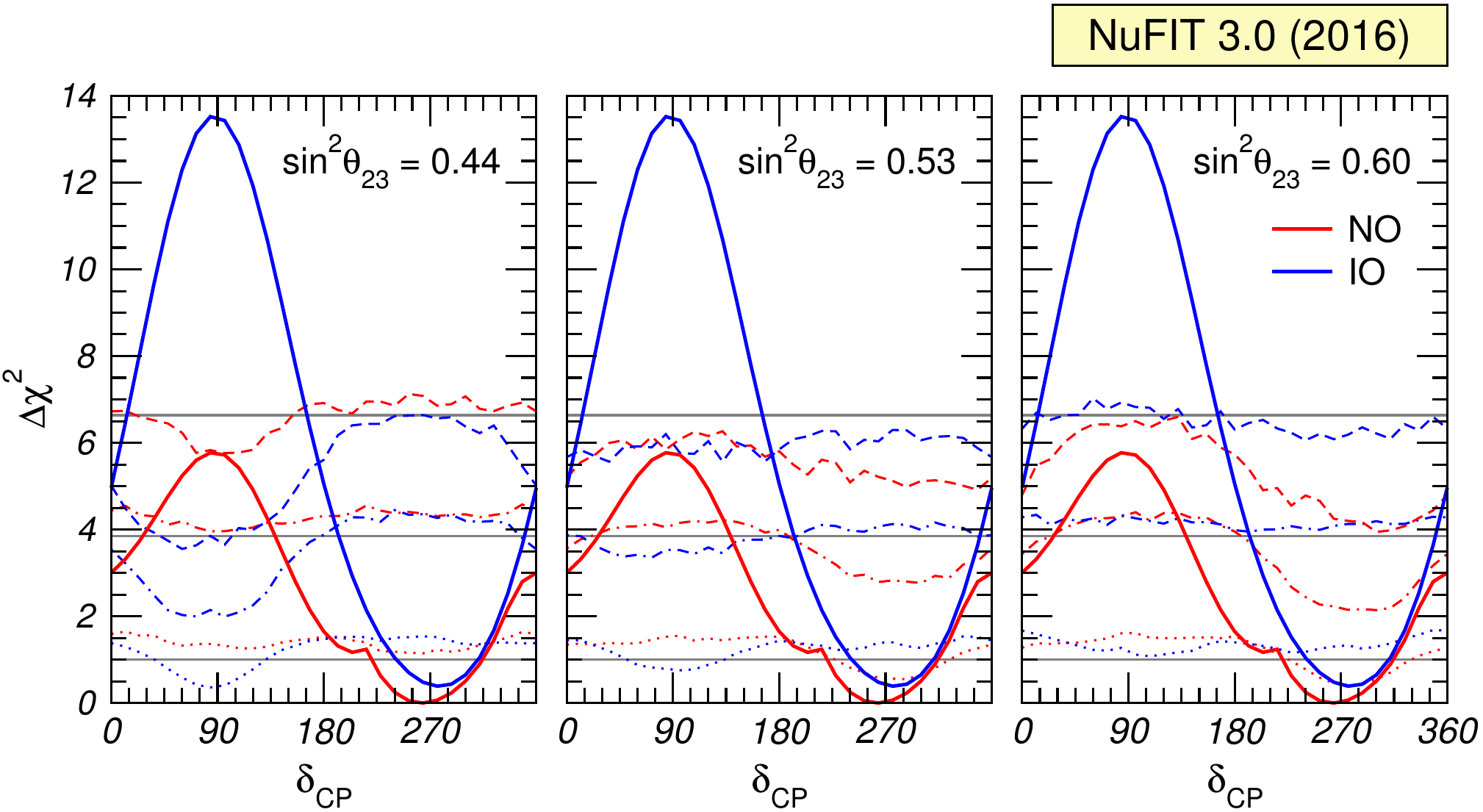}
  \caption{68\%, 95\% and 99\% confidence levels (broken curves) for
    the test statistics~\eqref{eq:testStatistics1} along with its
    value (solid curves) for the combination of T2K, NO$\nu$A, MINOS
    and reactor data.  The value of $\sin^2\theta_{23}$ given in each
    panel corresponds to the assumed true value chosen to generate the
    pseudo-experiments and for all panels we take
    $\Dmq_{3\ell,\text{true}} = -2.53 \times 10^{-3}~\text{eV}^2$ for
    IO and $+2.54 \times 10^{-3}~\text{eV}^2$ for NO.  The solid
    horizontal lines represent the 68\%, 95\% and 99\% CL predictions
    from Wilks' theorem.}
  \label{fig:probab-dcp}
\end{figure}

The value of the test statistics~\eqref{eq:testStatistics1} is shown
in Fig.~\ref{fig:probab-dcp} for the combination of T2K, NO$\nu$A,
MINOS and Daya-Bay as a function of $\delta_\text{CP}$ for both mass
orderings. In the generation of the pseudo-data we have assumed three
representative values of $\theta_{23,\text{true}}$ as shown in the
plots.  The broken curves show, for each set of true values, the values
of $\Delta\chi^2(\delta_\text{CP}, \text{O})$ which are larger than
68\%, 95\%, and 99\% of all generated data samples.

From the figure we read that if the $\Delta\chi^2$ from real data
(solid curve, identical in the three panels) for a given ordering is
above the $x\%$ CL lines for that ordering for a given value of
$\delta_\text{CP}$, that value of $\delta_\text{CP}$ and the mass
ordering can be rejected with $x\%$ confidence. So if the minimum of
the $\Delta\chi^2$ curve for one of the orderings (in this case IO is
the one with non-zero minimum) is above the $x\%$ CL line one infers
that that ordering is rejected at that CL.

For the sake of comparison we also show in Fig.~\ref{fig:probab-dcp}
the corresponding 68\%, 95\% and 99\% Gaussian confidence levels as
horizontal lines.  There are some qualitative deviations from
Gaussianity that have already been reported~\cite{Elevant:2015ska}:
\begin{itemize}
\item For $\theta_{23}<45^\circ$, $\delta_\text{CP} = 90^\circ$, and
  IO as well as for $\theta_{23} > 45^\circ$, $\delta_\text{CP} =
  270^\circ$ and NO, the confidence levels decrease. This effect
  arises because at those points in parameter space the $\nu_\mu \to
  \nu_e$ oscillation probability has a minimum or a maximum,
  respectively. Therefore, statistical fluctuations leading to less
  (or more) events than predicted cannot be accommodated by adjusting
  the parameters.  $\Delta \chi^2$ is small more often and the
  confidence levels decrease. This is an effect always present at
  boundaries in parameter space, usually referred to as an effective
  decrease in the number of degrees of freedom in the model.

\item Conversely for $\delta_\text{CP} \sim 90^\circ$ for
  $\theta_{23}>45^\circ$, and $\delta_\text{CP} \sim 270^\circ$ for
  $\theta_{23}<45^\circ$, the confidence levels increase. This is
  associated with the prominent presence of the octant
  degeneracy. Degeneracies imply that statistical fluctuations can
  drive you away from the true value, $\Delta \chi^2$ increases, and
  the confidence levels increase. This is usually referred to as an
  effective increase in the number of degrees of freedom in the model
  due to degeneracies.

\item Overall we find that with present data confidence levels are
  clearly closer to Gaussianity than found in
  Refs.~\cite{Gonzalez-Garcia:2014bfa, Elevant:2015ska}, where similar
  simulations have been performed with less data available. For those
  data sets confidence levels were consistently below their Gaussian
  limit. This was mainly a consequence of the limited statistics and
  the cyclic nature of $\delta_\text{CP}$ which lead to an effective
  decrease in the number of degrees of freedom.  We now find that when
  the full combination of data currently available is included this
  effect is reduced, as expected if experiments become more sensitive.

\item For all true values considered, IO is not rejected even at
  $1\sigma$.  In particular we find IO disfavored at $30\% - 40\%$ for
  $\sin^2\theta_{23}=0.44 - 0.60$.
\end{itemize}

\begin{table}\centering
  \begin{tabular}{cc|ccc}
    \hline\hline
    $\sin^2\theta_{23,\text{true}}$ & Ordering & CP cons.
    & 90\% CL range & 95\% CL range
    \\
    \hline
    0.44
    & NO & 70\%
    & $[0^\circ, 14^\circ] \cup [151^\circ, 360^\circ]$
    & $[0^\circ, 37^\circ] \cup [133^\circ, 360^\circ]$
    \\
    & IO & 98\%
    & $[200^\circ, 341^\circ]$
    & $[190^\circ, 350^\circ]$
    \\
    0.53
    & NO & 70\%
    & $[150^\circ, 342^\circ]$
    & $[0^\circ, 28^\circ] \cup [133^\circ, 360^\circ]$
    \\
    & IO & 98\%
    & $[203^\circ, 342^\circ]$
    & $[193^\circ, 350^\circ]$
    \\
    0.60
    & NO & 70\%
    & $[148^\circ, 336^\circ]$
    & $[0^\circ, 28^\circ] \cup [130^\circ, 360^\circ]$
    \\
    & IO & 97\%
    & $[205^\circ, 345^\circ]$
    & $[191^\circ, 350^\circ]$
    \\
    \hline
    Gaussian
    & NO & 80\%
    & $[158^\circ, 346^\circ]$
    & $[0^\circ, 26^\circ] \cup [139^\circ, 360^\circ]$
    \\
    & IO & 97\%
    & $[208^\circ, 332^\circ]$
    & $[193^\circ, 350^\circ]$
    \\
    \hline\hline
  \end{tabular}
  \caption{Confidence level with which CP conservation
    ($\delta_\text{CP} = 0, 180^\circ$) is rejected (third column) and
    90\% and 95\% confidence intervals for $\delta_\text{CP}$ (fourth
    and fifth column) for different sets of true values of the
    parameters and in the Gaussian approximation.  Confidence
    intervals for $\delta_\text{CP}$ as well as the CL for CP
    conservation are defined for both orderings with respect to the
    global minimum (which happens for NO).}
  \label{tab:CPCL}
\end{table}

Quantitatively we show in Tab.~\ref{tab:CPCL} the CL at which CP
conservation ($\delta_\text{CP}=0, 180^\circ$) is disfavored as well
as the 90\% and 95\% confidence intervals for $\delta_\text{CP}$.  We
find that the CL of rejection of CP conservation as well as the
allowed ranges do not depend very significantly on
$\theta_{23,\text{true}}$. This can be understood from
Fig.~\ref{fig:probab-dcp}: the dependence on
$\theta_{23,\text{true}}$ occur mostly for $\delta_\text{CP} \sim
90^\circ$ and IO, a region discarded with a large CL, and for
$\delta_\text{CP} \sim 270^\circ$ and NO, a region around the best
fit.

Note that in the table the intervals for $\delta_\text{CP}$ are
defined for both orderings with respect to the global minimum (which
happens for NO). Hence the intervals for IO include the effect that IO
is slightly disfavored with respect to NO. They cannot be directly
compared to the intervals given in Tab.~\ref{tab:bfranges}, where we
defined intervals relative to the local best fit point for each
ordering.

A similar comment applies also to the CL quoted in the table to reject
CP conservation. For IO this is defined relative to the best fit point
in NO. We find that for NO, CP conservation is allowed at 70\% CL,
\textit{i.e.}, slightly above $1\sigma$ (with some deviations from the
Gaussian result of 80\%~CL), while for IO the CL for CP conservation
is above $2\sigma$. Note that values of $\delta_\text{CP} \simeq
90^\circ$ are disfavored at around 99\%~CL for NO, while for IO the
rejection is at even higher CL: the $\Delta\chi^2$ with respect to the
global minimum is around 14, which would correspond to $3.7\sigma$ in
the Gaussian limit. Our Monte Carlo sample of $10^4$ pseudo-data sets
is not large enough to confirm such a high confidence level.

\subsection{$\theta_{23}$ and the mass ordering}

\begin{figure}\centering
  \includegraphics[width=\textwidth]{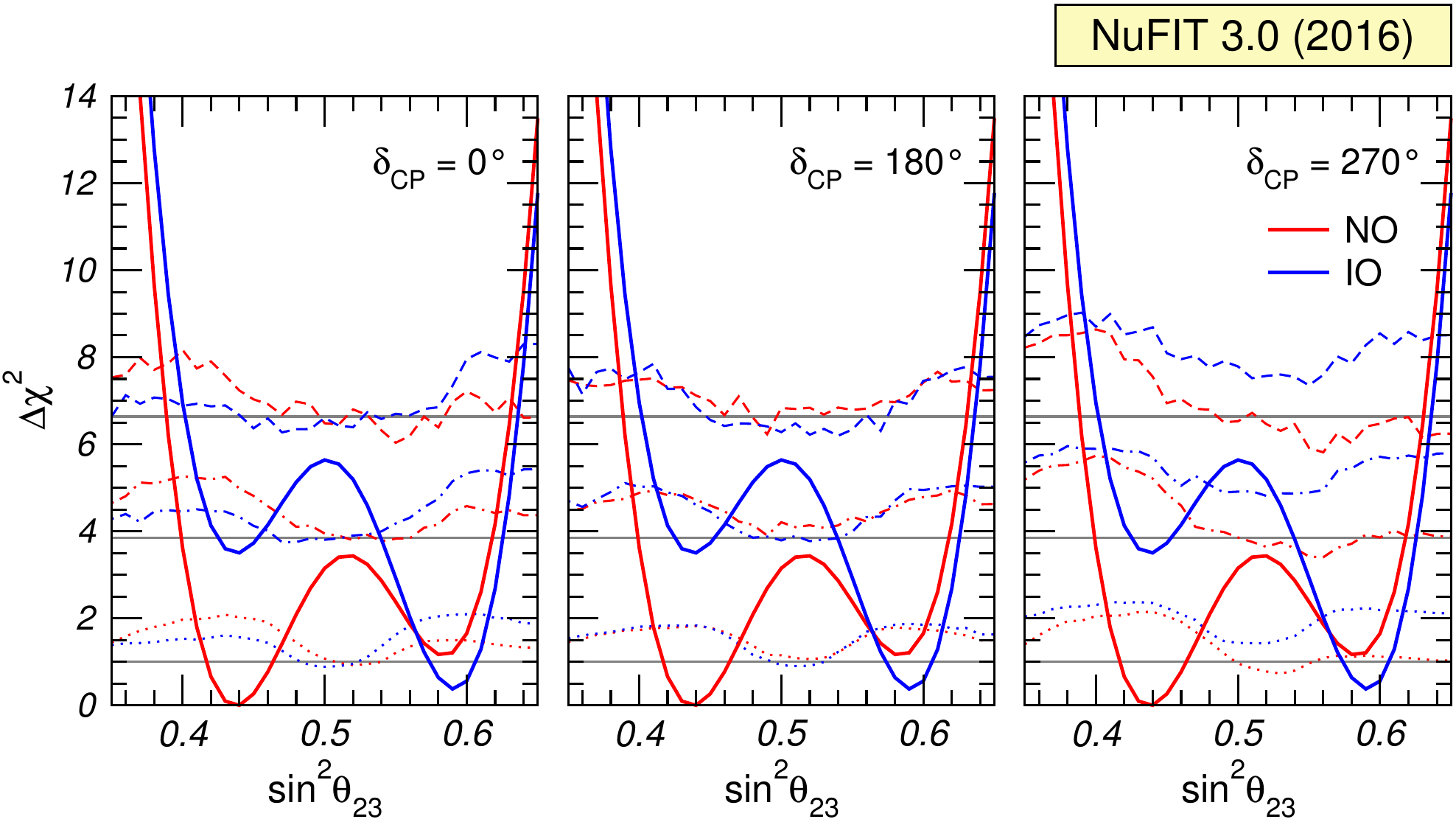}
  \caption{68\%, 95\% and 99\% confidence levels (broken curves) for
    the test statistics~\eqref{eq:testStatistics2} along with its
    value (solid curves) for the combination of T2K, NO$\nu$A, MINOS
    and reactor data.  The value of $\delta_\text{CP}$ above each plot
    corresponds to the assumed true value chosen to generate the
    pseudo-experiments and for all panels we take
    $\Dmq_{3\ell,\text{true}} = -2.53 \times 10^{-3}~\text{eV}^2$ for
    IO and $+2.54 \times 10^{-3}~\text{eV}^2$ for NO.  The solid
    horizontal lines represent the 68\%, 95\% and 99\% CL predictions
    from Wilks' theorem.}
  \label{fig:probab-t23}
\end{figure}

Moving now to the discussion of $\theta_{23}$, we show the value of
the test statistics~\eqref{eq:testStatistics2} in
Fig.~\ref{fig:probab-t23} for the combination of T2K, NO$\nu$A, MINOS
and Daya-Bay experiments as a function of $\theta_{23}$, for both mass
orderings.  For the generation of the pseudo-data we have assumed
three example values $\delta_\text{CP,true} = 0, 180^\circ,
270^\circ$. We do not show results for $\delta_\text{CP,true}
=90^\circ$, since this value is already quite disfavored by data,
especially for IO.\footnote{We are aware of the fact that this choice
  is somewhat arbitrary and implicitly resembles Bayesian
  reasoning. In the strict frequentist sense we cannot a priori
  exclude any true value of the parameters.}  The broken curves show
for each set of true values, the values of $\Delta\chi^2(\theta_{23},
\text{O})$ which are larger than 68\%, 95\%, and 99\% of all generated
data samples. From the figure we see that the deviations from
Gaussianity are not very prominent and can be understood as follows:
\begin{itemize}
\item The confidence levels decrease around maximal mixing because of
  the boundary on the parameter space present at maximal mixing for
  disappearance data.

\item There is some increase and decrease in the confidence levels for
  $\delta_\text{CP} = 270^\circ$, in the same parameter region as the
  corresponding ones in Fig.~\ref{fig:probab-dcp}.
\end{itemize}

\begin{table}\centering
  \catcode`?=\active\def?{\hphantom{0}}
  \begin{tabular}{cc|ccc}
    \hline\hline
    $\delta_\text{CP,true}$ & Ordering & $\theta_{23} = 45^\circ$
    & 90\% CL range & 95\% CL range
    \\
    \hline
    $??0^\circ$
    & NO & 92\%
    & $[0.40, 0.49] \cup [0.55, 0.61]$
    & $[0.39, 0.62]$
    \\
    & IO & 98\%
    & $[0.55, 0.62]$
    & $[0.42, 0.46] \cup [0.54, 0.63]$
    \\
    $180^\circ$
    & NO & 91\%
    & $[0.40, 0.50] \cup [0.54, 0.61]$
    & $[0.40, 0.62]$
    \\
    & IO & 98\%
    & $[0.43, 0.44] \cup [0.55, 0.62]$
    & $[0.41, 0.46] \cup [0.54, 0.63]$
    \\
    $270^\circ$
    & NO & 92\%
    & $[0.40, 0.49] \cup [0.55, 0.61]$
    & $[0.39, 0.62]$
    \\
    & IO & 97\%
    & $[0.42, 0.45] \cup [0.55, 0.62]$
    & $[0.41, 0.48] \cup [0.53, 0.63]$
    \\
    \hline
    Gaussian & NO & 92\%
    & $[0.41, 0.49] \cup [0.55, 0.61]$
    & $[0.40, 0.62]$
    \\
    & IO & 98\%
    & $[0.56, 0.62]$
    & $[0.43, 0.45] \cup [0.54, 0.63]$
    \\
    \hline\hline
  \end{tabular}
  \caption{CL for the rejection of maximal $\theta_{23}$ mixing (third
    column), and 90\% and 95\% CL intervals for $\sin^2\theta_{23}$
    for different sets of true parameter values and in the Gaussian
    approximation (last row).}
  \label{tab:t23CL}
\end{table}

In Tab.~\ref{tab:t23CL} we show the CL at which the combination of LBL
and reactor experiments can disfavor maximal $\theta_{23}$ mixing
($\theta_{23} = 45^\circ$) as well as the 90\% and 95\% confidence
intervals for $\sin^2\theta_{23}$ for both orderings with respect to
the global best fit. We observe from the table that the Gaussian
approximation is quite good for both, the CL of maximal mixing as well
as for the confidence intervals. We conclude that present data
excludes maximal mixing at slightly more than 90\%~CL. Again we note
that the intervals for $\sin^2\theta_{23}$ for IO cannot be directly
compared with the ones from Tab.~\ref{tab:bfranges}, where they are
defined with respect to the local minimum in each ordering.

\begin{table}\centering
  \catcode`?=\active\def?{\hphantom{0}}
  \begin{tabular}{c|ccc}
    \hline\hline
    $\delta_\text{CP,true}$ & NO/2nd Oct. & IO/1st Oct. & IO/2nd Oct.
    \\
    \hline
    $??0^\circ$ & 62\% & 91\% & 28\%
    \\
    $180^\circ$ & 56\% & 89\% & 32\%
    \\
    $270^\circ$ & 70\% & 83\% & 27\%
    \\
    \hline
    Gaussian & 72\% & 94\% & 46\%
    \\
    \hline\hline
  \end{tabular}
  \caption{CL for the rejection of various combinations of mass
    ordering and $\theta_{23}$ octant with respect to the global best
    fit (which happens for NO and 1st octant). We quote the CL of the
    local minima for each ordering/octant combination, assuming three
    example values for the true value of $\delta_\text{CP}$ as well as
    for the Gaussian approximation (last row).}
  \label{tab:octantCL}
\end{table}

In Tab.~\ref{tab:octantCL} we show the CL at which a certain
combination of mass ordering and $\theta_{23}$ octant can be excluded
with respect to the global minimum in the NO and 1st $\theta_{23}$
octant.  We observe that the CL of the second octant for NO shows
relatively large deviations from Gaussianity and dependence on the
true value of $\delta_\text{CP}$. In any case, the sensitivity is very
low and the 2nd octant can be reject at most at 70\% CL ($1\sigma$)
for all values of $\delta_\text{CP}$. The first octant for IO can be
excluded at between 83\% and 91\%~CL, depending on $\delta_\text{CP}$.
As discussed above, the exclusion of the IO/2nd octant case
corresponds also to the exclusion of the IO, since at that point the
confidence interval in IO would vanish. Also in this case we observe
deviations from the Gaussian approximation and the CL of at best 32\%
is clearly less than $1\sigma$ (consistent with the results discussed
in the previous subsection), showing that the considered data set has
essentially no sensitivity to the mass ordering.

\section{Conclusions}
\label{sec:summary}

We have presented the results of the updated (as of fall 2016)
analysis of relevant neutrino data in the framework of mixing among
three massive neutrinos.  Quantitatively the present determination of
the two mass differences, three mixing angles and the relevant CP
violating phase obtained under the assumption that their
log-likelihood follows a $\chi^2$ distribution is listed in
Table~\ref{tab:bfranges}, and the corresponding leptonic mixing matrix
is given in Eq.~\eqref{eq:umatrix}.  We have found that the maximum
allowed CP violation in the leptonic sector parametrized by the
Jarlskog determinant is $J_\text{CP}^\text{max} = 0.0329 \pm 0.0007 \,
(^{+0.0021}_{-0.0024}))$ at $1\sigma$ ($3\sigma$).

We have studied in detail how the sensitivity to the least-determined
parameters $\theta_{23}$, $\delta_\text{CP}$ and the mass ordering
depends on the proper combination of the different data samples
(Sec.~\ref{subsec:dm32}).  Furthermore we have quantified deviations
from the Gaussian approximation in the evaluation of the confidence
intervals for $\theta_{23}$ and $\delta_\text{CP}$ by performing a
Monte Carlo study of the long baseline accelerator and reactor results
(Sec.~\ref{sec:MC}).  We can summarize the main conclusions in these
sections as follows:
\begin{itemize}
\item At present the precision on the determination of
  $|\Dmq_{3\ell}|$ from $\nu_\mu$ disappearance in LBL accelerator
  experiments NO$\nu$A, T2K and MINOS is comparable to that from
  $\nu_e$ disappearance in reactor experiments, in particular with the
  spectral information from Daya-Bay. When comparing the region for
  each LBL experiment with that of the reactor experiments we find
  some dispersion in the best fit values and allowed ranges.

\item The interpretation of the data from accelerator LBL experiments
  in the framework of $3\nu$ mixing requires using information from
  the reactor experiments, in particular about the mixing angle
  $\theta_{13}$.  But since, as mentioned above, reactor data also
  constrain $|\Dmq_{3\ell}|$, the resulting CL of presently low
  confidence effects (in particular the non-maximality of
  $\theta_{23}$ and the mass ordering) is affected by the inclusion of
  this information in the combination.

\item We find that the mass ordering favored by NO$\nu$A changes from
  NO to IO when the information on $\Dmq_{3\ell}$ from reactor
  experiments is correctly included in the LBL+REA combination, and
  the $\Delta\chi^2$ of NO in T2K is reduced from around 2 to 0.5 (see
  Fig.~\ref{fig:chisq-dma}).  Our MC study of the combination of LBL
  and reactor data shows that for all cases generated, NO is favored
  but with a CL of less than $1\sigma$.

\item About the non-maximality of $\theta_{23}$, we find that when the
  information on $\Dmq_{3\ell}$ from reactor experiments is correctly
  included in the LBL+REA combination, it is not NO$\nu$A but actually
  MINOS which contributes most to the preference for non-maximal
  $\theta_{23}$ (see Fig.~\ref{fig:chisq-t23}).  Quantitatively our MC
  study of the combination of LBL and reactor data shows that for all
  the cases generated the CL for rejection of maximal $\theta_{23}$ is
  about 92\% for NO.  As seen in Fig.~\ref{fig:probab-t23} and
  Tab.~\ref{tab:t23CL}, the CL of maximal mixing as well as confidence
  intervals for $\sin^2\theta_{23}$ derived with MC simulations are
  not very different from the corresponding Gaussian approximation.

\item The same study shows that for NO (IO) the favored octant is
  $\theta_{23}<45^\circ$ ($\theta_{23}>45^\circ$).  The CL for
  rejection of the disfavored octant depends on the true value of
  $\delta_\text{CP}$ assumed in the MC study and it is generically
  lower than the one obtained in the Gaussian limit (see
  Tab.~\ref{tab:octantCL}). For example, for NO the second octant is
  disfavored at a confidence level between $0.9\sigma$ and $1.3\sigma$
  depending on the assumed true value of $\delta_\text{CP}$.

\item The present sensitivity to $\delta_\text{CP}$ is driven by T2K
  with a minor contribution from NO$\nu$A for IO (see
  Fig.~\ref{fig:chisq-dcp}). The dependence of the combined CL of the
  ``hint'' towards leptonic CP violation and in particular for
  $\delta_\text{CP} \simeq 270^\circ$ on the true value of
  $\theta_{23}$ is shown in Fig.~\ref{fig:probab-dcp}, from which we
  read that for all cases generated CP conservation is disfavored only
  at 70\% ($1.05\sigma$) for NO.  Values of $\delta_\text{CP} \simeq
  90^\circ$ are disfavored at around 99\%~CL for NO, while for IO the
  rejection is at higher CL ($\Delta\chi^2 \simeq 14$ with respect to
  the global minimum).
\end{itemize}
Finally we comment that the increased statistics in SK4 and Borexino
has had no major impact in the long-standing tension between the best
fit values of $\Dmq_{21}$ as determined from the analysis of KamLAND
and solar data, which remains an unresolved $\sim 2\sigma$ effect.

Future updates of this analysis will be provided at the NuFIT website
quoted in Ref.~\cite{nufit}.

\section*{Acknowledgments}

This work is supported by USA-NSF grant PHY-1620628, by EU Networks
FP10 ITN ELUSIVES (H2020-MSCA-ITN-2015-674896) and INVISIBLES-PLUS
(H2020-MSCA-RISE-2015-690575), by MINECO grants FPA2013-46570,
FPA2012-31880 and MINECO/FEDER-UE grant FPA2015-65929-P, by Maria de
Maetzu program grant MDM-2014-0367 of ICCUB, and the ``Severo Ochoa''
program grant SEV-2012-0249 of IFT. I.E.\ acknowledges support from
the FPU program fellowship FPU15/03697.

\appendix

\section{List of data used in the analysis}
\label{sec:appendix}

\section*{Solar experiments}

\begin{itemize}
\setlength{\itemsep}{0mm}
\item Chlorine total rate~\cite{Cleveland:1998nv}, 1 data point.

\item Gallex \& GNO total rates~\cite{Kaether:2010ag}, 2 data points.

\item SAGE total rate~\cite{Abdurashitov:2009tn}, 1 data point.

\item SK1 full energy and zenith spectrum~\cite{Hosaka:2005um}, 44
  data points.

\item SK2 full energy and day/night spectrum~\cite{Cravens:2008aa}, 33
  data points.

\item SK3 full energy and day/night spectrum~\cite{Abe:2010hy}, 42
  data points.

\item SK4 2055-day day-night asymmetry~\cite{sksol:nakano2016} and
  2365-day energy spectrum~\cite{sksol:ichep2016}, 24 data points.

\item SNO combined analysis~\cite{Aharmim:2011vm}, 7 data points.

\item Borexino Phase-I 740.7-day low-energy
  data~\cite{Bellini:2011rx}, 33 data points.

\item Borexino Phase-I 246-day high-energy data~\cite{Bellini:2008mr},
  6 data points.

\item Borexino Phase-II 408-day low-energy
  data~\cite{Bellini:2014uqa}, 42 data points.
\end{itemize}

\section*{Atmospheric experiments}

\begin{itemize}
\item IceCube/DeepCore 3-year data~\cite{Aartsen:2014yll,
  deepcore:2016}, 64 data points.
\end{itemize}

\section*{Reactor experiments}

\begin{itemize}
\setlength{\itemsep}{0mm}
\item KamLAND combined DS1 \& DS2 spectrum~\cite{Gando:2010aa}, 17
  data points.

\item CHOOZ energy spectrum~\cite{Apollonio:1999ae}, 14 data points.

\item Palo-Verde total rate~\cite{Piepke:2002ju}, 1 data point.

\item Double-Chooz FD-I (461 days) and FD-II (212 days)
  spectra~\cite{dc:moriond2016}, 54 data points.

\item Daya-Bay 1230-day spectrum~\cite{db:nu2016}, 34 data
  points.

\item Reno 800-day near \& far total rates~\cite{reno:nu2014}, 2
  data points (with free normalization).

\item SBL reactor data (including Daya-Bay total flux at near
  detector), 77 data points~\cite{Kopp:2013vaa, db:nu2014}.
\end{itemize}

\section*{Accelerator experiments}

\begin{itemize}
  \setlength{\itemsep}{0mm}
\item MINOS $10.71\times 10^{20}$~pot $\nu_\mu$-disappearance
  data~\cite{Adamson:2013whj}, 39 data points.

\item MINOS $3.36\times 10^{20}$~pot $\bar\nu_\mu$-disappearance
  data~\cite{Adamson:2013whj}, 14 data points.

\item MINOS $10.6\times 10^{20}$~pot $\nu_e$-appearance
  data~\cite{Adamson:2013ue}, 5 data points.

\item MINOS $3.3\times 10^{20}$~pot $\bar\nu_e$-appearance
  data~\cite{Adamson:2013ue}, 5 data points.

\item T2K $7.48\times 10^{20}$ pot $\nu_\mu$-disappearance
  data~\cite{t2k:ichep2016, t2k:susy2016}, 28 data points.

\item T2K $7.48\times 10^{20}$ pot $\nu_e$-appearance
  data~\cite{t2k:ichep2016, t2k:susy2016}, 5 data points.

\item T2K $7.47\times 10^{20}$ pot $\bar\nu_\mu$-disappearance
  data~\cite{t2k:ichep2016, t2k:susy2016}, 63 data points.

\item T2K $7.47\times 10^{20}$ pot $\bar\nu_e$-appearance
  data~\cite{t2k:ichep2016, t2k:susy2016}, 1 data point.

\item NO$\nu$A $6.05\times 10^{20}$ pot $\nu_\mu$-disappearance
  data~\cite{nova:nu2016}, 18 data points.

\item NO$\nu$A $6.05\times 10^{20}$ pot $\nu_e$-appearance
  data~\cite{nova:nu2016}, 10 data points.
\end{itemize}

\bibliographystyle{JHEP}
\bibliography{references}

\providecommand{\href}[2]{#2}\begingroup\raggedright\begin{thebibliography}{10}

\bibitem{Pontecorvo:1967fh}
B.~Pontecorvo, \emph{{Neutrino experiments and the question of leptonic-charge
  conservation}}, {\emph{Sov. Phys. JETP} {\bf 26} (1968) 984--988}.

\bibitem{Gribov:1968kq}
V.~N. Gribov and B.~Pontecorvo, \emph{{Neutrino astronomy and lepton charge}},
  \href{http://dx.doi.org/10.1016/0370-2693(69)90525-5}{\emph{Phys. Lett.} {\bf
  B28} (1969) 493}.

\bibitem{GonzalezGarcia:2007ib}
M.~C. Gonzalez-Garcia and M.~Maltoni, \emph{{Phenomenology with Massive
  Neutrinos}},
  \href{http://dx.doi.org/10.1016/j.physrep.2007.12.004}{\emph{Phys. Rept.}
  {\bf 460} (2008) 1--129}, [\href{http://arxiv.org/abs/0704.1800}{{\tt
  0704.1800}}].

\bibitem{Giunti:2015wnd}
C.~Giunti, \emph{{Light Sterile Neutrinos: Status and Perspectives}},
  \href{http://dx.doi.org/10.1016/j.nuclphysb.2016.01.013}{\emph{Nucl. Phys.}
  {\bf B908} (2016) 336--353}, [\href{http://arxiv.org/abs/1512.04758}{{\tt
  1512.04758}}].

\bibitem{Maki:1962mu}
Z.~Maki, M.~Nakagawa and S.~Sakata, \emph{{Remarks on the unified model of
  elementary particles}},
  \href{http://dx.doi.org/10.1143/PTP.28.870}{\emph{Prog. Theor. Phys.} {\bf
  28} (1962) 870--880}.

\bibitem{Kobayashi:1973fv}
M.~Kobayashi and T.~Maskawa, \emph{{CP Violation in the Renormalizable Theory
  of Weak Interaction}},
  \href{http://dx.doi.org/10.1143/PTP.49.652}{\emph{Prog. Theor. Phys.} {\bf
  49} (1973) 652--657}.

\bibitem{Bilenky:1980cx}
S.~M. Bilenky, J.~Hosek and S.~T. Petcov, \emph{{On Oscillations of Neutrinos
  with Dirac and Majorana Masses}},
  \href{http://dx.doi.org/10.1016/0370-2693(80)90927-2}{\emph{Phys. Lett.} {\bf
  B94} (1980) 495}.

\bibitem{Langacker:1986jv}
P.~Langacker, S.~T. Petcov, G.~Steigman and S.~Toshev, \emph{{On the
  Mikheev-Smirnov-Wolfenstein (MSW) Mechanism of Amplification of Neutrino
  Oscillations in Matter}},
  \href{http://dx.doi.org/10.1016/0550-3213(87)90699-7}{\emph{Nucl. Phys.} {\bf
  B282} (1987) 589}.

\bibitem{Gonzalez-Garcia:2014bfa}
M.~C. Gonzalez-Garcia, M.~Maltoni and T.~Schwetz, \emph{{Updated fit to three
  neutrino mixing: status of leptonic CP violation}},
  \href{http://dx.doi.org/10.1007/JHEP11(2014)052}{\emph{JHEP} {\bf 11} (2014)
  052}, [\href{http://arxiv.org/abs/1409.5439}{{\tt 1409.5439}}].

\bibitem{Capozzi:2016rtj}
F.~Capozzi, E.~Lisi, A.~Marrone, D.~Montanino and A.~Palazzo, \emph{{Neutrino
  masses and mixings: Status of known and unknown $3\nu$ parameters}},
  \href{http://dx.doi.org/10.1016/j.nuclphysb.2016.02.016}{\emph{Nucl. Phys.}
  {\bf B908} (2016) 218--234}, [\href{http://arxiv.org/abs/1601.07777}{{\tt
  1601.07777}}].

\bibitem{Forero:2014bxa}
D.~V. Forero, M.~Tortola and J.~W.~F. Valle, \emph{{Neutrino oscillations
  refitted}}, \href{http://dx.doi.org/10.1103/PhysRevD.90.093006}{\emph{Phys.
  Rev.} {\bf D90} (2014) 093006}, [\href{http://arxiv.org/abs/1405.7540}{{\tt
  1405.7540}}].

\bibitem{Minakata:2002jv}
H.~Minakata, H.~Sugiyama, O.~Yasuda, K.~Inoue and F.~Suekane, \emph{{Reactor
  measurement of theta(13) and its complementarity to long baseline
  experiments}}, \href{http://dx.doi.org/10.1103/PhysRevD.70.059901,
  10.1103/PhysRevD.68.033017}{\emph{Phys. Rev.} {\bf D68} (2003) 033017},
  [\href{http://arxiv.org/abs/hep-ph/0211111}{{\tt hep-ph/0211111}}]. [Erratum:
  Phys. Rev.D70,059901(2004)].

\bibitem{Huber:2003pm}
P.~Huber, M.~Lindner, T.~Schwetz and W.~Winter, \emph{{Reactor neutrino
  experiments compared to superbeams}},
  \href{http://dx.doi.org/10.1016/S0550-3213(03)00493-0}{\emph{Nucl. Phys.}
  {\bf B665} (2003) 487--519}, [\href{http://arxiv.org/abs/hep-ph/0303232}{{\tt
  hep-ph/0303232}}].

\bibitem{Huber:2004ug}
P.~Huber, M.~Lindner, M.~Rolinec, T.~Schwetz and W.~Winter, \emph{{Prospects of
  accelerator and reactor neutrino oscillation experiments for the coming ten
  years}}, \href{http://dx.doi.org/10.1103/PhysRevD.70.073014}{\emph{Phys.
  Rev.} {\bf D70} (2004) 073014},
  [\href{http://arxiv.org/abs/hep-ph/0403068}{{\tt hep-ph/0403068}}].

\bibitem{Cleveland:1998nv}
B.~T. Cleveland et~al., \emph{{Measurement of the solar electron neutrino flux
  with the Homestake chlorine detector}},
  \href{http://dx.doi.org/10.1086/305343}{\emph{Astrophys. J.} {\bf 496} (1998)
  505--526}.

\bibitem{Kaether:2010ag}
F.~Kaether, W.~Hampel, G.~Heusser, J.~Kiko and T.~Kirsten, \emph{{Reanalysis of
  the GALLEX solar neutrino flux and source experiments}},
  \href{http://dx.doi.org/10.1016/j.physletb.2010.01.030}{\emph{Phys. Lett.}
  {\bf B685} (2010) 47--54}, [\href{http://arxiv.org/abs/1001.2731}{{\tt
  1001.2731}}].

\bibitem{Abdurashitov:2009tn}
{\scshape SAGE} collaboration, J.~N. Abdurashitov et~al., \emph{{Measurement of
  the solar neutrino capture rate with gallium metal. III: Results for the
  2002--2007 data-taking period}},
  \href{http://dx.doi.org/10.1103/PhysRevC.80.015807}{\emph{Phys. Rev.} {\bf
  C80} (2009) 015807}, [\href{http://arxiv.org/abs/0901.2200}{{\tt
  0901.2200}}].

\bibitem{Hosaka:2005um}
{\scshape Super-Kamiokande} collaboration, J.~Hosaka et~al., \emph{{Solar
  neutrino measurements in Super-Kamiokande-I}},
  \href{http://dx.doi.org/10.1103/PhysRevD.73.112001}{\emph{Phys. Rev.} {\bf
  D73} (2006) 112001}, [\href{http://arxiv.org/abs/hep-ex/0508053}{{\tt
  hep-ex/0508053}}].

\bibitem{Cravens:2008aa}
{\scshape Super-Kamiokande} collaboration, J.~Cravens et~al., \emph{{Solar
  neutrino measurements in Super-Kamiokande-II}},
  \href{http://dx.doi.org/10.1103/PhysRevD.78.032002}{\emph{Phys. Rev.} {\bf
  D78} (2008) 032002}, [\href{http://arxiv.org/abs/0803.4312}{{\tt
  0803.4312}}].

\bibitem{Abe:2010hy}
{\scshape Super-Kamiokande} collaboration, K.~Abe et~al., \emph{{Solar neutrino
  results in Super-Kamiokande-III}},
  \href{http://dx.doi.org/10.1103/PhysRevD.83.052010}{\emph{Phys. Rev.} {\bf
  D83} (2011) 052010}, [\href{http://arxiv.org/abs/1010.0118}{{\tt
  1010.0118}}].

\bibitem{sksol:nakano2016}
Y.~Nakano, \emph{{$^8$B solar neutrino spectrum measurement using
  Super-Kamiokande IV}}.
\newblock PhD thesis, Tokyo U., 2016-02.

\bibitem{sksol:ichep2016}
Y.~Nakano, ``{Solar neutrino results from Super-Kamiokande}.'' Talk given at
  the {\it 38th International Conference on High Energy Physics}, Chicago, USA,
  August 3--10, 2016.

\bibitem{Aharmim:2011vm}
{\scshape SNO} collaboration, B.~Aharmim et~al., \emph{{Combined Analysis of
  all Three Phases of Solar Neutrino Data from the Sudbury Neutrino
  Observatory}},
  \href{http://dx.doi.org/10.1103/PhysRevC.88.025501}{\emph{Phys. Rev.} {\bf
  C88} (2013) 025501}, [\href{http://arxiv.org/abs/1109.0763}{{\tt
  1109.0763}}].

\bibitem{Bellini:2011rx}
{\scshape Borexino} collaboration, G.~Bellini et~al., \emph{{Precision
  measurement of the 7Be solar neutrino interaction rate in Borexino}},
  \href{http://dx.doi.org/10.1103/PhysRevLett.107.141302}{\emph{Phys. Rev.
  Lett.} {\bf 107} (2011) 141302}, [\href{http://arxiv.org/abs/1104.1816}{{\tt
  1104.1816}}].

\bibitem{Bellini:2008mr}
{\scshape Borexino} collaboration, G.~Bellini et~al., \emph{{Measurement of the
  solar 8B neutrino rate with a liquid scintillator target and 3 MeV energy
  threshold in the Borexino detector}},
  \href{http://dx.doi.org/10.1103/PhysRevD.82.033006}{\emph{Phys. Rev.} {\bf
  D82} (2010) 033006}, [\href{http://arxiv.org/abs/0808.2868}{{\tt
  0808.2868}}].

\bibitem{Bellini:2014uqa}
{\scshape BOREXINO} collaboration, G.~Bellini et~al., \emph{{Neutrinos from the
  primary proton–proton fusion process in the Sun}},
  \href{http://dx.doi.org/10.1038/nature13702}{\emph{Nature} {\bf 512} (2014)
  383--386}.

\bibitem{Adamson:2013whj}
{\scshape MINOS} collaboration, P.~Adamson et~al., \emph{{Measurement of
  Neutrino and Antineutrino Oscillations Using Beam and Atmospheric Data in
  MINOS}}, \href{http://dx.doi.org/10.1103/PhysRevLett.110.251801}{\emph{Phys.
  Rev. Lett.} {\bf 110} (2013) 251801},
  [\href{http://arxiv.org/abs/1304.6335}{{\tt 1304.6335}}].

\bibitem{Adamson:2013ue}
{\scshape MINOS} collaboration, P.~Adamson et~al., \emph{{Electron neutrino and
  antineutrino appearance in the full MINOS data sample}},
  \href{http://dx.doi.org/10.1103/PhysRevLett.110.171801}{\emph{Phys. Rev.
  Lett.} (2013) }, [\href{http://arxiv.org/abs/1301.4581}{{\tt 1301.4581}}].

\bibitem{t2k:ichep2016}
K.~Iwamoto, ``{Recent Results from T2K and Future Prospects}.'' Talk given at
  the {\it 38th International Conference on High Energy Physics}, Chicago, USA,
  August 3--10, 2016.

\bibitem{t2k:susy2016}
A.~Cervera, ``{Latest Results from Neutrino Oscillation Experiments}.'' Talk
  given at the SUSY~2016 Conference, Melbourne, Australia, July 3--8, 2016.

\bibitem{nova:nu2016}
P.~Vahle, ``{New results from NOvA}.'' Talk given at the {\it XXVII
  International Conference on Neutrino Physics and Astrophysics}, London, UK,
  July 4--9, 2016.

\bibitem{Gando:2010aa}
{\scshape KamLAND} collaboration, A.~Gando et~al., \emph{{Constraints on
  $\theta_{13}$ from A Three-Flavor Oscillation Analysis of Reactor
  Antineutrinos at KamLAND}},
  \href{http://dx.doi.org/10.1103/PhysRevD.83.052002}{\emph{Phys. Rev.} {\bf
  D83} (2011) 052002}, [\href{http://arxiv.org/abs/1009.4771}{{\tt
  1009.4771}}].

\bibitem{Apollonio:1999ae}
{\scshape CHOOZ} collaboration, M.~Apollonio et~al., \emph{{Limits on Neutrino
  Oscillations from the CHOOZ Experiment}},
  \href{http://dx.doi.org/10.1016/S0370-2693(99)01072-2}{\emph{Phys. Lett.}
  {\bf B466} (1999) 415--430}, [\href{http://arxiv.org/abs/hep-ex/9907037}{{\tt
  hep-ex/9907037}}].

\bibitem{Piepke:2002ju}
{\scshape Palo Verde} collaboration, A.~Piepke, \emph{{Final results from the
  Palo Verde neutrino oscillation experiment}},
  \href{http://dx.doi.org/10.1016/S0146-6410(02)00117-5}{\emph{Prog. Part.
  Nucl. Phys.} {\bf 48} (2002) 113--121}.

\bibitem{dc:moriond2016}
M.~Ishitsuka, ``{New results of Double Chooz}.'' {Talk given at the Conference
  {\it Rencontres de Moriond EW 2016}, La Thuile, Italy, March 12--19, 2016}.

\bibitem{db:nu2016}
Z.~Yu, ``{Recent Results from the Daya Bay Experiment}.'' Talk given at the
  {\it XXVII International Conference on Neutrino Physics and Astrophysics},
  London, UK, July 4--9, 2016.

\bibitem{reno:nu2014}
S.-H. Seo, ``{New Results from RENO}.'' Talk given at the {\it XXVI
  International Conference on Neutrino Physics and Astrophysics}, Boston, USA,
  June 2--7, 2014.

\bibitem{Kopp:2013vaa}
J.~Kopp, P.~A.~N. Machado, M.~Maltoni and T.~Schwetz, \emph{{Sterile Neutrino
  Oscillations: The Global Picture}},
  \href{http://dx.doi.org/10.1007/JHEP05(2013)050}{\emph{JHEP} {\bf 1305}
  (2013) 050}, [\href{http://arxiv.org/abs/1303.3011}{{\tt 1303.3011}}].

\bibitem{Kwon:1981ua}
H.~Kwon, F.~Boehm, A.~Hahn, H.~Henrikson, J.~Vuilleumier et~al., \emph{Search
  for neutrino oscillations at a fission reactor},
  \href{http://dx.doi.org/10.1103/PhysRevD.24.1097}{\emph{Phys.Rev.} {\bf D24}
  (1981) 1097--1111}.

\bibitem{Zacek:1986cu}
{\scshape CALTECH-SIN-TUM} collaboration, G.~Zacek et~al., \emph{{Neutrino
  Oscillation Experiments at the Gosgen Nuclear Power Reactor}},
  \href{http://dx.doi.org/10.1103/PhysRevD.34.2621}{\emph{Phys.Rev.} {\bf D34}
  (1986) 2621--2636}.

\bibitem{Vidyakin:1987ue}
G.~Vidyakin, V.~Vyrodov, I.~Gurevich, Y.~Kozlov, V.~Martemyanov et~al.,
  \emph{Detection of anti-neutrinos in the flux from two reactors},
  {\emph{Sov.Phys.JETP} {\bf 66} (1987) 243--247}.

\bibitem{Vidyakin:1994ut}
G.~Vidyakin, V.~Vyrodov, Y.~Kozlov, A.~Martemyanov, V.~Martemyanov et~al.,
  \emph{{Limitations on the characteristics of neutrino oscillations}},
  {\emph{JETP Lett.} {\bf 59} (1994) 390--393}.

\bibitem{Afonin:1988gx}
A.~Afonin, S.~Ketov, V.~Kopeikin, L.~Mikaelyan, M.~Skorokhvatov et~al.,
  \emph{{A study of the reaction $\bar\nu_e + p \to e^+ + n$ on a nuclear
  reactor}}, {\emph{Sov.Phys.JETP} {\bf 67} (1988) 213--221}.

\bibitem{Kuvshinnikov:1990ry}
A.~Kuvshinnikov, L.~Mikaelyan, S.~Nikolaev, M.~Skorokhvatov and A.~Etenko,
  \emph{{Measuring the $\bar\nu_e + p \to n + e^+$ cross-section and beta decay
  axial constant in a new experiment at Rovno NPP reactor. (In Russian)}},
  {\emph{JETP Lett.} {\bf 54} (1991) 253--257}.

\bibitem{Declais:1994su}
Y.~Declais, J.~Favier, A.~Metref, H.~Pessard, B.~Achkar et~al., \emph{{Search
  for neutrino oscillations at 15-meters, 40-meters, and 95-meters from a
  nuclear power reactor at Bugey}},
  \href{http://dx.doi.org/10.1016/0550-3213(94)00513-E}{\emph{Nucl.Phys.} {\bf
  B434} (1995) 503--534}.

\bibitem{Declais:1994ma}
Y.~Declais, H.~de~Kerret, B.~Lefievre, M.~Obolensky, A.~Etenko et~al.,
  \emph{{Study of reactor anti-neutrino interaction with proton at Bugey
  nuclear power plant}},
  \href{http://dx.doi.org/10.1016/0370-2693(94)91394-3}{\emph{Phys.Lett.} {\bf
  B338} (1994) 383--389}.

\bibitem{Greenwood:1996pb}
Z.~D. Greenwood et~al., \emph{{Results of a two position reactor neutrino
  oscillation experiment}},
  \href{http://dx.doi.org/10.1103/PhysRevD.53.6054}{\emph{Phys. Rev.} {\bf D53}
  (1996) 6054--6064}.

\bibitem{Aartsen:2014yll}
{\scshape IceCube} collaboration, M.~Aartsen et~al., \emph{{Determining
  neutrino oscillation parameters from atmospheric muon neutrino disappearance
  with three years of IceCube DeepCore data}},
  \href{http://dx.doi.org/10.1103/PhysRevD.91.072004}{\emph{Phys. Rev.} {\bf
  D91} (2015) 072004}, [\href{http://arxiv.org/abs/1410.7227}{{\tt
  1410.7227}}].

\bibitem{nufit}
NuFIT webpage, \href{http://www.nu-fit.org}{\tt http://www.nu-fit.org}.

\bibitem{GonzalezGarcia:2012sz}
M.~Gonzalez-Garcia, M.~Maltoni, J.~Salvado and T.~Schwetz, \emph{{Global fit to
  three neutrino mixing: critical look at present precision}},
  \href{http://dx.doi.org/10.1007/JHEP12(2012)123}{\emph{JHEP} {\bf 1212}
  (2012) 123}, [\href{http://arxiv.org/abs/1209.3023}{{\tt 1209.3023}}].

\bibitem{Mueller:2011nm}
T.~Mueller, D.~Lhuillier, M.~Fallot, A.~Letourneau, S.~Cormon et~al.,
  \emph{{Improved Predictions of Reactor Antineutrino Spectra}},
  \href{http://dx.doi.org/10.1103/PhysRevC.83.054615}{\emph{Phys.Rev.} {\bf
  C83} (2011) 054615}, [\href{http://arxiv.org/abs/1101.2663}{{\tt
  1101.2663}}].

\bibitem{Huber:2011wv}
P.~Huber, \emph{{On the determination of anti-neutrino spectra from nuclear
  reactors}}, \href{http://dx.doi.org/10.1103/PhysRevC.85.029901,
  10.1103/PhysRevC.84.024617}{\emph{Phys.Rev.} {\bf C84} (2011) 024617},
  [\href{http://arxiv.org/abs/1106.0687}{{\tt 1106.0687}}].

\bibitem{Mention:2011rk}
G.~Mention, M.~Fechner, T.~Lasserre, T.~Mueller, D.~Lhuillier et~al.,
  \emph{{The Reactor Antineutrino Anomaly}},
  \href{http://dx.doi.org/10.1103/PhysRevD.83.073006}{\emph{Phys.Rev.} {\bf
  D83} (2011) 073006}, [\href{http://arxiv.org/abs/1101.2755}{{\tt
  1101.2755}}].

\bibitem{Schwetz:2006md}
T.~Schwetz, \emph{{What is the probability that theta(13) and CP violation will
  be discovered in future neutrino oscillation experiments?}},
  \href{http://dx.doi.org/10.1016/j.physletb.2007.02.053}{\emph{Phys. Lett.}
  {\bf B648} (2007) 54--59}, [\href{http://arxiv.org/abs/hep-ph/0612223}{{\tt
  hep-ph/0612223}}].

\bibitem{Blennow:2014sja}
M.~Blennow, P.~Coloma and E.~Fernandez-Martinez, \emph{{Reassessing the
  sensitivity to leptonic CP violation}},
  \href{http://dx.doi.org/10.1007/JHEP03(2015)005}{\emph{JHEP} {\bf 03} (2015)
  005}, [\href{http://arxiv.org/abs/1407.3274}{{\tt 1407.3274}}].

\bibitem{GonzalezGarcia:2003qf}
M.~C. Gonzalez-Garcia and C.~Pena-Garay, \emph{{Three neutrino mixing after the
  first results from K2K and KamLAND}},
  \href{http://dx.doi.org/10.1103/PhysRevD.68.093003}{\emph{Phys. Rev.} {\bf
  D68} (2003) 093003}, [\href{http://arxiv.org/abs/hep-ph/0306001}{{\tt
  hep-ph/0306001}}].

\bibitem{Jarlskog:1985ht}
C.~Jarlskog, \emph{{Commutator of the Quark Mass Matrices in the Standard
  Electroweak Model and a Measure of Maximal CP Violation}},
  \href{http://dx.doi.org/10.1103/PhysRevLett.55.1039}{\emph{Phys.Rev.Lett.}
  {\bf 55} (1985) 1039}.

\bibitem{PDG}
K.~A. Olive, \emph{{Review of Particle Physics}},
  \href{http://dx.doi.org/10.1088/1674-1137/40/10/100001}{\emph{Chin. Phys.}
  {\bf C40} (2016) 100001}.

\bibitem{Wolfenstein:1977ue}
L.~Wolfenstein, \emph{{Neutrino oscillations in matter}},
  \href{http://dx.doi.org/10.1103/PhysRevD.17.2369}{\emph{Phys. Rev.} {\bf D17}
  (1978) 2369--2374}.

\bibitem{Mikheev:1986gs}
S.~P. Mikheev and A.~Y. Smirnov, \emph{{Resonance enhancement of oscillations
  in matter and solar neutrino spectroscopy}}, {\emph{Sov. J. Nucl. Phys.} {\bf
  42} (1985) 913--917}.

\bibitem{Bergstrom:2016cbh}
J.~Bergstrom, M.~C. Gonzalez-Garcia, M.~Maltoni, C.~Pena-Garay, A.~M. Serenelli
  and N.~Song, \emph{{Updated determination of the solar neutrino fluxes from
  solar neutrino data}},
  \href{http://dx.doi.org/10.1007/JHEP03(2016)132}{\emph{JHEP} {\bf 03} (2016)
  132}, [\href{http://arxiv.org/abs/1601.00972}{{\tt 1601.00972}}].

\bibitem{Vinyoles:2016djt}
N.~Vinyoles, A.~M. Serenelli, F.~L. Villante, S.~Basu, J.~Bergström, M.~C.
  Gonzalez-Garcia et~al., \emph{{A new Generation of Standard Solar Models}},
  \href{http://arxiv.org/abs/1611.09867}{{\tt 1611.09867}}.

\bibitem{Bezerra:2012at}
T.~J.~C. Bezerra, H.~Furuta and F.~Suekane, \emph{{Measurement of Effective
  $\Delta m_{31}^2$ using Baseline Differences of Daya Bay, RENO and Double
  Chooz Reactor Neutrino Experiments}},
  \href{http://arxiv.org/abs/1206.6017}{{\tt 1206.6017}}.

\bibitem{Seo:2016uom}
H.~Seo et~al., \emph{{Spectral Measurement of the Electron Antineutrino
  Oscillation Amplitude and Frequency using 500 Live Days of RENO Data}},
  \href{http://arxiv.org/abs/1610.04326}{{\tt 1610.04326}}.

\bibitem{Wendell:2014dka}
{\scshape Super-Kamiokande} collaboration, R.~Wendell, \emph{{Atmospheric
  Results from Super-Kamiokande}},
  \href{http://dx.doi.org/10.1063/1.4915569}{\emph{AIP Conf. Proc.} {\bf 1666}
  (2015) 100001}, [\href{http://arxiv.org/abs/1412.5234}{{\tt 1412.5234}}].

\bibitem{skatm:nufact2016}
J.~Kameda, ``{Recent results from Super-Kamokande on atmospheric neutrinos and
  next project: Hyper-Kamioande}.'' Talk given at the {\it XII Rencontres de
  Vietnam: NuFact 2016}, Qui Nhon, Vietnam, August 21--27, 2016.

\bibitem{skatm:thesis}
K.~P. Lee, ``{Study of the neutrino mass hierarchy with the atmospheric
  neutrino data observed in SuperKamiokande}.'' {Ph.D. thesis, The University
  of Tokyo}, 2012.

\bibitem{Elevant:2015ska}
J.~Elevant and T.~Schwetz, \emph{{On the determination of the leptonic CP
  phase}}, \href{http://dx.doi.org/10.1007/JHEP09(2015)016}{\emph{JHEP} {\bf
  09} (2015) 016}, [\href{http://arxiv.org/abs/1506.07685}{{\tt 1506.07685}}].

\bibitem{Blennow:2013oma}
M.~Blennow, P.~Coloma, P.~Huber and T.~Schwetz, \emph{{Quantifying the
  sensitivity of oscillation experiments to the neutrino mass ordering}},
  \href{http://dx.doi.org/10.1007/JHEP03(2014)028}{\emph{JHEP} {\bf 03} (2014)
  028}, [\href{http://arxiv.org/abs/1311.1822}{{\tt 1311.1822}}].

\bibitem{Wilks:1938dza}
S.~S. Wilks, \emph{{The Large-Sample Distribution of the Likelihood Ratio for
  Testing Composite Hypotheses}},
  \href{http://dx.doi.org/10.1214/aoms/1177732360}{\emph{Annals Math. Statist.}
  {\bf 9} (1938) 60--62}.

\bibitem{deepcore:2016}
{\scshape IceCube} collaboration, J.~P. Yañez et~al., ``{IceCube Oscillations:
  3 years muon neutrino disappearance data}.''
  \href{http://icecube.wisc.edu/science/data/nu_osc}{\tt
  http://icecube.wisc.edu/science/data/nu\_osc}.

\bibitem{db:nu2014}
C.~Zhang, ``{Recent Results From Daya Bay}.'' Talk given at the {\it XXVI
  International Conference on Neutrino Physics and Astrophysics}, Boston, USA,
  June 2--7, 2014.

\end{thebibliography}\endgroup

\end{document}